\documentclass[12pt,preprint]{aastex}
\usepackage{graphicx}
\usepackage{appendix}
\usepackage{lscape}
\pdfoutput=1

\slugcomment{resubmitted 25 Nov 2010}

\begin{document}{}

\def\t0{\theta_{\circ}}
\def\muo{\mu_{\circ}}
\def\sd{\partial}
\def\be{\begin{equation}}
\def\en{\end{equation}}
\def\bv{\bf v}
\def\bvo{\bf v_{\circ}}
\def\ro{r_{\circ}}
\def\rhoo{\rho_{\circ}}
\def\etal{et al.\ }
\def\msun{\,M_{\sun}}
\def\rsun{\,R_{\sun}}
\def\rstar{\,R_{star}}
\def\lsun{L_{\sun}}
\def\mbol{M_{bol}}
\def\msunyr{M_{\sun} yr^{-1}}
\def\kms{\rm \, km \, s^{-1}}
\def\mdot{\dot{M}}
\def\mdotd{\dot{M}_{\rm d}}
\def\Md{\dot{M}}
\def\curf{{\cal F}}
\def\ecs{erg cm^{-2} s^{-1}}
\def \haebe{HAeBe}
\def \mum {\,{\rm \mu m}}
\def \simali {{\sim\,}}
\def \K {\,{\rm K}}
\def \Angstrom     {\,{\rm \AA}}
\newcommand \g            {\,{\rm g}}
\newcommand{\ltapp}{\raisebox{-.4ex}{\rlap{$\sim$}} \raisebox{.4 ex}{$<$}}
\newcommand{\gtapp}{\raisebox{-.4ex}{\rlap{$\sim$}} \raisebox{.4 ex}{$>$}}
\newcommand \cm           {\,{\rm cm}}

\title{The Mass-Loss Return from Evolved Stars to the Large Magellanic 
Cloud IV: Construction and Validation of a Grid of Models for Oxygen-Rich 
AGB Stars, Red Supergiants, and Extreme AGB Stars}

\author{Benjamin A. Sargent\altaffilmark{1},
S. Srinivasan\altaffilmark{2},
M. Meixner\altaffilmark{1}
}

\altaffiltext{1}{Space Telescope Science Institute, 3700 San Martin 
                 Drive, Baltimore, MD 21218, USA;
                 {\sf sargent@stsci.edu}}
\altaffiltext{2}{Institut d'Astrophysique de Paris, 98 bis, Boulevard Arago, 
                 Paris 75014, France}

\begin{abstract}

To measure the mass loss from dusty oxygen-rich (O-rich) evolved stars 
in the Large Magellanic Cloud (LMC), we have constructed a grid of 
models of spherically-symmetric dust shells around stars with constant 
mass-loss rates using {\bf 2D}ust.  These models will constitute the O-rich 
model part of the ``Grid of Red supergiant and Asymptotic giant branch star 
ModelS'' (GRAMS).  This model grid explores 4 
parameters -- stellar effective temperature from 2100\,K--4700\,K; luminosity 
from 10$^{3}$--10$^{6} \lsun$; dust shell inner radii of 3, 7, 11, 
and 15$\rstar$; and 10.0$\mum$ optical depth from 10$^{-4}$ to 26.  
From an initial grid of $\simali$1200 {\bf 2D}ust models, we create a larger 
grid of $\simali$69\,000 models by scaling to cover the luminosity range 
required by the data.  These models are offered to the public on a website.  
The matching in color-magnitude diagrams and color-color 
diagrams to observed O-rich asymptotic giant branch (AGB) and red supergiant 
(RSG) candidate stars from the SAGE and SAGE-Spec LMC samples and a 
small sample of OH/IR stars is generally very good.  The extreme AGB star 
candidates from SAGE are more consistent with carbon-rich (C-rich) than O-rich 
dust composition.  Our model grid suggests lower limits to the mid-infrared colors 
of the dustiest AGB stars for which the chemistry could be O-rich.  Finally, the 
fitting of GRAMS models to SEDs of sources fit by other studies 
provides additional verification of our grid and anticipates future, more 
expansive efforts.

\end{abstract}

\keywords{circumstellar matter, infrared: stars, 
         stars: asymptotic giant branch}

\section{Introduction}

Near the ends of their lives, low- to intermediate-mass ($\ltapp$ 
8 $\msun$) stars become asymptotic giant branch (AGB) stars.  
``Asymptotic giant branch'' denotes the set of stars 
that approach the red giant stars asymptotically from the blue side 
in color, in color-magnitude diagrams of globular clusters \citep[e.g., 
see the analysis of M92 by][]{sandwal66}.  However, AGB stars 
reach higher luminosities than red giant stars, as revealed by their 
lower magnitudes than red giants at near-infrared wavelengths 
\citep[e.g.,][]{nw00,cioni00,blum06}.  In the Hertzsprung-Russell 
(HR) diagram, AGB stars may span a range of effective temperatures 
of 2000 to 5000\,K, depending upon such factors as stellar mass, 
metallicity, carbon-to-oxygen abundance ratio, etc.  \citet{groen09} 
find AGB stars' effective temperatures to span the range 2500--3800\,K.  
\citet{busso07} find carbon-rich AGB stars with effective temperatures 
as low as 2000\,K, while \citet{marig08} predict AGB stars of very low 
metallicity (Z $\simali$0.0001) can have effective temperatures as high 
as $\simali$5000\,K.  AGB stars' luminosities are typically $>$10$^{3} 
\lsun$ \citep[e.g.,][]{groen09}.

Stars in the AGB phase expel their own material.  Exactly how this 
mass-loss occurs is debated, but it is believed that radial pulsations 
lift material above the surface of the star, dust forms from this material, 
and the radiation pressure of photons from the star pushing on the 
newly-formed dust grains propels the dust away from the star, dragging 
the gas with it 
\citep{bowen88,flei91,flei92,feuch93,hoedo97,hoef98,wint97,wint00,jeo03,hoef09}.  
As a star enters 
further into its AGB phase, it increases its mass-loss 
\citep{ferrga06,srin09,groen09}.  Red supergiant (RSG) 
stars are massive stars ($>$ 8 $\msun$), that similarly eject their 
own material toward the ends 
of their lives.  This occurs as they evolve away from the Main 
Sequence and begin core helium burning, before they explode 
as supernovae \citep[see, e.g.,][]{mas03,verhoe09}.  As with 
AGB stars, RSG stars produce dust by a mechanism that is not 
yet totally understood, but this mechanism may be similar to 
that proposed for AGBs \citep{verhoe09}.  For a wide range of 
stellar masses, the end of a star's life is characterized by the 
return of matter back to the interstellar medium (ISM) from which 
the star formed originally.

The obscuration of much of the rest of our own Galaxy by its 
own ISM poses difficulties for studying evolved stars' production 
of dust and its subsequent life-cycle in the Milky Way.  For 
this reason, we turn to other galaxies for studying the life-cycle 
of dust and other matter.  The low average reddening of 
E(B-V)$\simali$ 0.075 \citep{schlegel98} toward the Large 
Magellanic Cloud (LMC), its favorable orientation \citep[24\arcdeg; 
see][]{weni01}, and especially its proximity \citep[$\simali$50kpc; 
see][]{fea99} ease determination of important stellar properties, 
such as luminosity (or, alternatively, absolute magnitudes).  This 
makes the LMC an optimal target for studying the life-cycle of dust 
in a galaxy.  This is the overarching theme for the Surveying the 
Agents of a Galaxy's Evolution (SAGE) {\it Spitzer} Space 
Telescope \citep{wer04} Legacy project \citep{meix06}.  The 
{\it Spitzer} Infrared Array Camera \citep[IRAC;][]{faz04} 
observations conducted as part of the SAGE project found over 
6 million stars in the LMC, including thousands of oxygen-rich 
(O-rich) AGB stars, carbon-rich (C-rich) AGB stars, and extreme AGB 
stars \citep{blum06,srin09}; RSGs \citep{bonan09}; and Young Stellar 
Objects \citep{whit08,grchu09} 
in various stages of evolution.  By using the infrared emission from 
dust, we can study the life cycle of matter as it is expelled from evolved 
stars, stored in the ISM of the LMC, gathered together in 
star-forming regions, and incorporated into YSOs to form 
planetary systems around new stars.

Spectral Energy Distributions (SEDs), plots of emitted power versus 
wavelength, are used to constrain various properties of AGB 
stars.  One means by which AGB properties are constrained by their 
SEDs is to fit models to the SEDs of individual AGB stars 
\citep[e.g.,][]{vl99,groen09,sarg10,srin10}.  
Another approach, perhaps better suited to constraining general 
properties of relatively large samples of these stars, is to compute a 
grid of models and compare tracks or grids of these 
models to data in color-magnitude diagrams (CMDs; plots of magnitude 
versus color) or color-color diagrams (CCDs; plots of one color 
versus another).  For instance, \citet{vokw88} and \citet{marig08} 
compute radiative transfer models of AGB stars that also take into 
account the evolution of these stars, and one of the methods by 
which they evaluate their resultant evolutionary model tracks is to 
compare them to observed data in CCDs \citep{vokw88} and CMDs 
\citep{marig08}.  Others more simply use CMDs and CCDs to verify 
their radiative transfer model tracks or grids 
\citep[e.g.,]{ivez95,groen06,gonzlo10}, though \citet{groen06} also 
simulate evolution from the AGB to post-AGB stages.  We adopt the 
approach of constructing a model grid and comparing 
it to observed data in CMDs and CCDs \citep[e.g.,][]{ivez95,gonzlo10}, 
though we do not attempt to model evolution within the AGB stage or 
from the AGB to the post-AGB stages as others have done.  In 
subsequent papers, we will return to the first approach and fit 
the SED of each AGB and RSG candidates in the SAGE database 
individually.

In this paper we present the grid of models we will later use to 
study the mass loss of those stars in the SAGE 
sample identified as candidate O-rich AGB stars, extreme AGB stars, and RSG 
stars.  Using the dust properties found to be reasonable to use in 
radiative transfer (RT) modeling of the SEDs of two O-rich AGB stars 
by \citet{sarg10}, we have computed a grid of thousands of {\bf 2D}ust 
\citep{um03} radiative transfer models for O-rich evolved stars.  {\bf 2D}ust 
computes radiative transfer through an axisymmetric dust envelope 
surrounding a star.  It accomplishes this by dividing the envelope into a 
finite number of cells and computing the radiative transfer through each 
cell.  In a companion paper, Srinivasan et al. (2010, {\it in prep}, ``Paper V'') 
are creating a similar model grid for C-rich AGB stars.  We test our 
O-rich model grid by comparing in color-magnitude 
diagrams (CMDs) and color-color diagrams (CCDs) the colors and 
magnitudes of our models to those of observed O-rich AGB, 
RSG, and extreme AGB candidates identified in the SAGE 
sample.  In anticipation of the next step of using our model grid to fit the SED 
of each candidate O-rich AGB, RSG, and extreme AGB in the 
SAGE sample, we also include sample fits in this paper to stars whose 
SEDs have been modeled in other studies, as a test of our model grid.

\section{Model Grid}

In order to model the thousands of oxygen-rich and extreme AGB star 
candidates and $\simali$100 RSG candidates, we have constructed a 
grid of $\simali$69\,000 {\bf 2D}ust models.  For a summary of the 
model parameters explored in our grid that will be discussed in this 
section, see Table 1.

\subsection{Initial Grid}

To create the final grid of 68\,600 models, we first created an 
initial grid of 1225 models, which we discuss in this subsection.  The 
final grid explores a range of values in four parameters that have the 
greatest effect upon the output model spectra.  These four parameters 
are stellar effective temperature (T$_{\rm eff}$), dust shell inner radius 
(R$_{\rm min}$), optical depth at 10.0$\mum$ ($\tau_{\rm 10}$), and 
stellar luminosity (L$_{\rm star}$).  The first three are true input 
parameters, as they are required to be chosen before a model can be 
run (this includes T$_{\rm eff}$, as the choice of this determines which 
stellar photosphere model is used in the model).  These three are 
inputs for each {\bf 2D}ust model in the initial grid, and a range of values 
for each of them is explored in the initial grid.  As will be discussed 
individually in detail later, we explored 14 values of T$_{\rm eff}$, 4 
values of R$_{\rm min}$, and 39 values of $\tau_{\rm 10}$.  This gives a 
total of 2184 total possible models in the initial grid, but of these, only 
1225 were kept because the rest had the unphysical condition of dust 
hotter than the dust sublimation temperature at the dust shell inner radius.  
The last parameter, luminosity, is technically a calculated output 
parameter (as we discuss later, it is obtained by integrating the input stellar 
photosphere model 
over all wavelengths, then using distance to the star to obtain luminosity; 
dust mass-loss rate is also a calculated output) from each model.  
In constructing the final grid of models from the initial grid, {\it via} scaling 
relations that are described below, we do, however, explore a range of 56 
values of luminosity.  As all models in the final grid result from scaling 
models in the initial grid, all resultant scaled models were acceptable, 
so this gave a total of 1225$\times$56 = 68\,600 models in the final grid.  
The output model spectra and SEDs are then the true outputs of our 
models.  In addition to being sensitive to the four parameters we explore 
here, the output models are also sensitive to the properties assumed for 
the dust grains; however, these have already been explored by 
\citet{sarg10}.  The output spectra are less sensitive to the model 
parameters besides the grain properties and the four mentioned previously.  
Each model is computed using the techniques, procedures, and 
assumptions described by \citet{sarg10}.  Here, we enumerate and 
briefly discuss each of the various properties that serve as input to the 
code.

\subsubsection{Varied Model Parameters}

As noted previously, we explore both stellar photosphere properties 
and circumstellar dust shell properties in our model grid.  Here we 
summarize how we allowed these to vary.

\subsubsubsection{Stellar Properties}

To represent the stellar photosphere emission in each of our models, 
we use PHOENIX models \citep{kucin05,kucin06} for stars of 
subsolar metallicity (log({\it Z}/$Z_{\rm Sun}$) = -0.5) to match 
determinations of the metallicity of the LMC \citep[${\rm Z}_{\rm 
LMC}\simeq 0.5\times {\rm Z}_\odot$; see][]{duf82,bern08}.  PHOENIX 
models are given in increments of log({\it Z}/$Z_{\rm Sun}$) of 0.5, so 
the one we chose was nearest the LMC's metallicity.  We use PHOENIX 
models with stellar effective temperatures, T$_{\rm eff}$, between 2100\,K 
and 4700\,K, inclusive, in increments of 200\,K.  We briefly explored using 
PHOENIX models over a range of log(g/[cm*s$^{-2}$]) values from -0.5 to 
+2.5, but we found that this did not appreciably affect the colors of the 
output models, so we use PHOENIX models of log(g) = -0.5 only.

To determine the 
radius of a star for a single model, the mass of the star was assumed to be 
1 $\msun$, as this was the only stellar mass for which PHOENIX stellar 
photosphere models were available.  The effect of varying stellar mass 
was explored by \citet{kucin05}.  They found the optical-to-infrared colors 
{\it V}-{\it I} and {\it V}-{\it K} to be most affected by increasing the mass, with 
the colors becoming bluer by $\simali$0.1 magnitude when mass was 
increased from 1 $\msun$ to 5 $\msun$for a star with T$_{\rm eff}$ = 3500\,K 
and log(g) = 0.5.  The {\it B}-{\it V} and {\it J}-{\it K} colors were less affected, 
and all colors were less affected for hotter stars.  As many of the models in 
our grid have T$_{\rm eff} <$ 3500\,K, the optical-to-infrared colors of the 
naked stellar photospheres could be a few tenths of a magnitude bluer 
than the 1 $\msun$ ones whose models we use.  Greater realism in modeling 
stellar photosphere emission will have to await coverage of higher stellar 
masses in stellar photosphere model grids.

For all models in the initial grid, this 
constant combination of 1$\msun$ and log(g) = -0.5 gives a stellar radius 
of $\simali$295$\rsun$ (this follows not from the blackbody law but instead 
as a direct consequence of g = Mr$^{-2}$).  Since PHOENIX models are 
given in surface flux units, the choice of T$_{\rm eff}$ and log(g) = -0.5 
determines the surface flux of the star.  The distance to the LMC and the 
stellar radius then determines the flux density for the stellar photosphere model.  
From this flux, {\bf 2D}ust computes the stellar luminosity.  This combination of 
stellar radius of $\simali$295$\rsun$ and the stellar photosphere models we 
use (with a range of T$_{\rm eff}$ between 2100\,K and 4700\,K), in turn, gives 
a range of luminosities in the initial grid of 1500--38000 $\lsun$.  As we will 
describe later, for the final grid we explored a range of luminosities between 
10$^{3} \lsun$ and 10$^{6} \lsun$ using scaling relations on this initial grid.

\subsubsubsection{Shell Properties}

One property of the circumstellar dust shells that greatly affects the 
output models is optical depth at 10$\mum$, $\tau_{\rm 10}$.  As 
$\tau_{\rm 10}$ increases, the infrared excess increases.  This is 
expected from analytical arguments \citep[see Eq. 2 of][]{groen06}.  Our 
O-rich AGB, RSG, and extreme AGB sample spans a range from barely 
any excess above stellar photosphere emission to very large infrared 
excesses; therefore we allowed $\tau_{\rm 10}$ for our models to range 
from 10$^{-4}$ to 26.

The other property that had considerable effect on the output models was 
dust shell inner radius, R$_{\rm min}$.  This parameter determines the 
temperature of the hottest dust, which, in turn, determines the wavelength 
in the SED at which the emission from the dust shell becomes significant 
compared to that of the stellar photosphere.  The values of R$_{\rm min}$ 
explored in our grid were 3, 7, 11, and 15 $\times\rstar$.  We chose this 
range of R$_{\rm min}$ to correspond roughly to the allowable range of 
dust shell inner radii for O-rich AGB stars (2.5-14$\rstar$) according to 
\citet{hoef07}.  Note that the dust shell inner radii used by \citet{vl99} in 
modeling M stars are distributed over this range, and even in some cases 
approach 14$\rstar$.  The temperature of the hottest dust for a given model 
was allowed to be at most 1400\,K, as this is approximately the sublimation 
temperature of the astrophysically relevant silicates forsterite and enstatite 
\citep{posch07}.  We note that we do not have a model for every possible 
combination of T$_{\rm eff}$, luminosity, $\tau_{\rm 10}$, and R$_{\rm min}$, 
as models with their hottest dust higher than 1400\,K were rejected from 
our model grid.

\subsubsection{Fixed Properties}

Here we summarize those model parameters that are set at fixed values.  
Since we intend to keep the models as simple as possible, we explore 
only the model parameters to which the output spectra are most sensitive -- 
stellar effective temperature, stellar luminosity, dust shell optical depth, and 
dust shell inner radius.  We fix the rest of the parameters.  This includes 
those parameters relating to dust grain properties, which for this oxygen-rich 
model grid are exactly those used by \citet{sarg10}.  This also includes 
parameters relating to the dust shell geometry, as we assume for simplicity 
spherically-symmetric dust shells and constant mass-loss rate \citep[again, 
see][]{sarg10}.

\subsubsubsection{Dust Grains}

The models are sensitive to the choice of dust grain properties 
assumed.  These dust grain properties were explored by \citet{sarg10}, so we do not 
explore changing the dust properties here.  Instead, we briefly 
summarize the findings by \citet{sarg10}.  As a starting point for 
modeling the mass loss of evolved stars in the LMC, the SEDs 
of two O-rich AGBs have been modeled \citep{sarg10}.  From 
the ``faint'' population identified by \citet{blum06}, SSTISAGE1C 
J052206.92-715017.6 (hereafter, SSTSAGE052206) was chosen, 
and from the brighter population, HV 5715 was chosen.  Using 
{\bf 2D}ust \citep{um03}, RT models were constructed to fit 
the optical ({\it U} through {\it I}), near-infrared ({\it J} through {\it L}), 
IRAC, and Multiband Imaging 
Photometer for {\it Spitzer} \citep[MIPS;][]{riek04} broadband 
photometry and {\it Spitzer} Infrared Spectrograph 
\citep[IRS;][]{houck04} spectra (5--37$\mum$ wavelength) in 
the spectral energy distribution (SED) of each of 
these sources.  From this modeling, it was 
found that the complex indices of refraction of oxygen-deficient 
silicates from \citet{oss92} provided reasonable fits to both the 
overall photometry over all wavelengths probed and to the 
detailed IRS spectrum showing the silicate emission features 
at $\simali$10 and $\simali$20$\mum$.  The size distribution 
of these grains was assumed to be a KMH-like ``Power-law 
with Exponential Decay (PED)'' \citep{kmh94} distribution; that 
is, the number of grains of radius a, n(a), is proportional to 
$a^{\gamma}$e$^{-a/a_{\rm 0}}$, with $\gamma$ equal to -3.5, 
$a_{\rm min}$ of 0.01$\mum$, and $a_{\rm 0}$ of 0.1$\mum$.  
The models fit the data well, so we assume the dust properties 
used by \citet{sarg10} for the dust in the shells in our model grid.  
Mie theory is used to compute the absorption 
cross-sections of the grains.  These cross-sections are averaged with 
Harrington averaging \citep{har88} mode \citep[for a description, 
see][]{sarg10}.  These average cross-sections are then used in 
the {\bf 2D}ust radiative transfer calculations.  We assume isotropic 
scattering of radiation by the dust grains.

\subsubsubsection{Circumstellar Shell}

Spherically-symmetric mass-loss constant over time is assumed for our 
models, which means the density distribution of dust between the inner 
radius (R$_{\rm min}$) and outer radius (R$_{\rm max}$) goes as 
1/R$^{2}$.  For each model, R$_{\rm max}$ is set to be 
1000$\times$R$_{\rm min}$.  The dust expansion velocity, v$_{exp}$, 
for the dust shells is set at 10 km/s, and the distance assumed for our 
evolved star models is 50kpc \citep[for more, see][]{sarg10}.

\subsection{Expansion to the Full Grid}

The highest luminosity in our initial grid, $\simali$ 38000$\lsun$, is 
lower than most of the proposed lower limits on luminosity for a red 
supergiant \citep[see discussion on HV 5715 by][]{sarg10}.  To model 
RSGs and the AGB stars of higher luminosity than allowed by our initial 
grid, we use scaling relations on this grid.  For each model in the final 
grid, the scalar to be used is determined by dividing the luminosity of 
the model in the final grid by the luminosity of the model to be scaled 
from the initial grid.  We then multiply the output flux density at all 
wavelengths for the appropriate model from the initial grid by this scalar 
\citep{groen06}.  The radius of the star in the final grid is then determined 
by multiplying the stellar radius of the model in the initial grid by the square 
root of the scalar because L = 4$\pi$R$^{2}\sigma$T$^{4}$, and T does not 
change.  Further, the dust (and, therefore, total) mass-loss rate 
of the star is also scaled by the square root of the scalar \citep[this follows 
from Eq. 2 of][]{groen06}.  The other model parameters and outputs of 
concern do not change during this scaling.  The new grid has 41 values 
of luminosity between 10$^{3} \lsun$ and 2$\times$10$^{4} \lsun$, inclusive, 
with equal logarithmic spacing in luminosity, such that consecutive 
luminosity values differ by a factor of about 1.08.  The new grid also 
has 16 values of luminosity between 2$\times$10$^{4} \lsun$ and 
10$^{6} \lsun$, inclusive, with equal logarithmic spacing in luminosity, such 
that consecutive luminosity values differ by a factor of about 1.3.  This 
gives a total of 41 + 15 = 56 unique luminosity values in the new grid.

\subsection{Dust Mass-Loss Rate versus Luminosity}

In Figure 1 we plot dust mass-loss rate versus luminosity for the grid 
models that we plot in Figure 6.  Here we show both the initial grid (blue 
points) and final grid (black points), though in all the following figures we 
only show the final grid.  This illustrates the range of luminosities and 
dust mass-loss rates probed by our models.  The luminosities, as noted 
previously, range from 10$^{3}$ to $10^{6}$ $\lsun$.  At the lowest luminosities, 
the dust mass-loss rates probed is 3.0$\times$10$^{-13} \msunyr$ to 
1.0$\times$10$^{-6} \msunyr$.  At the highest luminosities, this range is 
1.0$\times$10$^{-11} \msunyr$ to 3.0$\times$10$^{-5} \msunyr$.

As we noted previously, we designed our grid to explore AGB star and 
dust shell parameters over large ranges of values, to be safe in allowing 
for modeling of anomalous, outlying stars in our sample.  In determining 
the mass return from evolved stars in the LMC, we are interested in the 
dust mass loss from stars of all mass-loss rates.  We note the 
spread in dust mass-loss rates in our model grid spans a range slightly 
larger than the range inferred from observations of AGB, RSG and red 
giant stars and predicted by theory for these stars.  At high dust mass-loss 
rates, our grid borders the dust mass-loss rate of $\simali$2.6$\times$10$^{-5} 
\msunyr$ at the luminosity of 2.2$\times$10$^{5} \lsun$ inferred by 
\citet{boyer10} for the RSG star IRAS 05280-6910.  At low dust mass-loss 
rates, our grid explores slightly lower values than those inferred from 
predictions by \citet{gail09} for early AGB (EAGB) stars.  The models of 
\citet{gail09} suggest the lowest total mass-loss rates for EAGB stars to be 
a few times 10$^{-10} \msun$, so, assuming a gas-to-dust mass ratio of a 
few hundred \citep[e.g., see][]{srin09,sarg10}, this implies dust mass-loss 
rates for EAGB stars of $\simali$10$^{-12} \msun$.  In support of the EAGB 
modeling by \citet{gail09}, these authors also predict total mass-loss rates 
for red giant stars similar to those for EAGB stars.  Again assuming a 
gas-to-dust mass ratio of a few hundred, one may infer the lowest dust 
mass-loss rates for red giant stars to be similar to those \citet{gail09} 
predict for EAGB stars, $\simali$10$^{-12} \msun$.  This is consistent with 
the lowest dust mass-loss rates inferred from observations of red giant stars 
in the globular cluster $\omega$ Centauri by \citet{mcdon09}.

\section{Computing the Grid}

A handful of test models was run on a MacBook 
Pro laptop computer with a 2.4 GHz Intel Core 2 Duo processor, 4 GB 
667 MHz DDR2 SDRAM memory, and Mac OS X version 10.5.8 
operating system.  These initial models determined useful values of 
model parameters like MXSTEP and VSPACE \citep[see][]{um03} that 
affect the performance and quality of the models\footnote{see also the 
User's Manual at http://www.stsci.edu/science/2dust/reference.cgi}.  
The parameter VSPACE must be as large as possible so that the 
integration step size for a radiative transfer calculation along a path is 
as small as possible (smaller than the local mean free path of photons); 
otherwise, the radiative transfer in the models will 
not be valid.  However, if VSPACE is too large, MXSTEP must increase 
to allow for a large number of very small steps in radiative transfer 
calculations.  The amount of virtual memory required for a model 
increases as MXSTEP increases, so it is good if MXSTEP is not too 
large.  The amount of virtual memory allowed for a single model 
calculation is just over 2 GB.  As we had a wide range of optical depths, 
we took a piecewise approach to the values of VSPACE/MXSTEP we 
used.  For different ranges of optical depth we chose different combinations 
of VSPACE and MXSTEP.  We verified for our models that the requirements 
were met of not only maximum integration step size (as set by VSPACE) 
not exceeding the local photon mean free path but also the maximum 
number of steps along a path not exceeding MXSTEP.

The models for our initial grid were run on the Royal Linux computing 
cluster at the Space Telescope Science Institute (STScI).  The cluster 
consists of 23 computing nodes, each with 2$\times$2$\times$2.4 
GHz processors, 8 GB RAM, 160 GB local disk space, and 17 TB of total 
disk space.  Royal enables up to 88 jobs to be computed simultaneously 
at high computing speeds and automates the scheduling of computing 
jobs.  Scripts written by the user enable the hundreds or thousands of 
models in the grid to be submitted for scheduling automatically, 
minimizing the amount of user intervention required to begin the 
computing of the grid.  Only the initial grid needed to be computed with 
Royal, as scaling required that no additional {\bf 2D}ust models to 
be run, only that existing {\bf 2D}ust models be scaled.  {\bf 2D}ust on 
Royal was run in non-interactive mode for these models, as it would not 
be able to run automatically in interactive mode on the Royal cluster.  
The number of radial grid points, NRAD, in the model shells was set to 
45, and the number of azimuthal grid points, NQ, in the model shells 
was set to 6.

It would be too time-consuming to try to run all the models desired 
serially on a single computer.  Though we ran a few test case models 
on a laptop computer, we made use of the Royal Linux computing 
cluster at Space Telescope Science Institute (STScI) to finish 
computation of our model grid in a reasonable amount of time.  Our 
initial grid had 4 values of R$_{\rm min}$, 14 values 
of T$_{eff}$, and 39 values of $\tau_{\rm 10}$, for a grand total of 
2\,184 models computed in our initial grid.  The Royal cluster can 
work on 88 different computing jobs separately at a given time.  For all 
models with $\tau_{\rm 10} <$ 1, the computation time per model was 
about 5 minutes, so the total processing time (the total time it would 
take to compute all models if only one model could be run at a time) 
was about 4$\times$14$\times$13$\times$(5 min) = 60.7 hours.  
Assuming 88 models could be computed simultaneously, this amounted 
to about 0.7 hours wall time (the total time elapsed while models were 
running, assuming 88 at a time, from when the first model began to 
when the last model finished).  For models with intermediate values of 
$\tau_{\rm 10}$ of 1--10, the computation time per model was about 1 
hour.  So for all models of $\tau_{\rm 10}$ between 1 and 10, the total 
processor time was about 4$\times$14$\times$10$\times$(1 hour) = 
560 hours, while the total wall time (again assuming 88 models could 
be computed at a time) was $\simali$6.4 hours.  For the rest of the 
models (those with $\tau_{\rm 10}$ between 11 and 26, the computation 
time per model was on average about 2 hours.    For all these, the total 
processor time was 4$\times$14$\times$16$\times$(2 hours) = 896 
hours, while the total wall time was $\simali$10.2 hours.  Altogether, the 
total processor time was $\simali$1500 hours, while the total wall time 
was about 0.7+6.4+10.2 = 17.3 hours.  Of the 2\,184 models that were 
computed in the initial grid, 1225 ($\simali$56\%) of them were kept and 
the others rejected because the dust temperature at R$_{\rm min}$ was 
higher than 1400\,K.  This initial grid was scaled to arrive at a final grid 
of 68\,600 models.

We call our model grid, to which we will add models for C-rich AGB stars 
(Srinivasan et al, {\it in prep}), the ``Grid of Red supergiant 
and Asymptotic giant branch star ModelS'' (GRAMS).  
We offer our final model grid to the public on a publicly-accessible 
website\footnote{see http://www.stsci.edu/science/2dust/grams\_models.cgi}.  Users 
can download the model {\bf 2D} spectra and files containing the input 
parameters and output values of 
interest, including dust mass-loss rate, dust temperature at dust shell 
inner radius, and the parameters listed in Table 1.  In addition, 
photometry synthesized from our {\bf 2D}ust models are also 
available on the website.  For details on the synthesis of these fluxes, 
see \citet{sarg10}.  Such synthesized fluxes can be transformed to 
magnitudes and compared to observed data on color-magnitude 
diagrams and color-color diagrams, as we do in the following section.

\section{Discussion}

We test the general usefulness of our model grid by comparing, in 
color-magnitude diagrams (CMDs) and color-color diagrams (CCDs), 
photometry synthesized from our grid to that observed for various 
evolved stars.  These evolved stars include SAGE sample candidate 
O-rich and extreme AGB stars and RSGs and O-rich AGB stars and 
RSGs in the SAGE-Spec sample.  Although the C-rich AGB stars are 
being modeled by Srinivasan et al. (2010, {\it in prep}; ``Paper V''), 
we include the C-rich AGB star sample from SAGE and SAGE-Spec to 
determine how well the models differentiate between the types, especially 
for the extreme AGB stars.  We identify the O-rich, C-rich, and extreme 
AGB star samples from SAGE that we include in our analyses by using the 
criteria given by \citet{srin09}.  We include RSGs from the SAGE sample as 
identified by \citet{bonan09}.  Also, we include O-rich AGB stars, RSGs, 
and C-rich AGB stars from SAGE-Spec that were identified as such by 
\citet{woods10}, using {\it Spitzer}-IRS spectra and ancillary photometry.  
For the SAGE-Spec sample, a star was only included here if it had a full 
5--37$\mum$ spectrum (as opposed to a 5--14$\mum$ spectrum only), 
in order that its dust chemistry be highly constrained.  
Some of the O-rich and C-rich AGB stars from SAGE-Spec are quite 
dusty and provide useful guidance in identifying the dust chemistry for 
the extreme AGB stars.  From SAGE-Spec, in total we include 34 O-rich 
AGB stars, 18 RSGs, and 47 C-rich AGB stars.  Additionally, we found 
SAGE photometry for the 10 OH/IR stars detected in OH maser emission 
by \citet{marsh04} and listed in their Table 2.  These OH/IR stars have 
properties (i.e., colors and magnitudes) that we might expect for O-rich 
extreme AGB stars.  As such, these stars represent an advanced stage 
of mass-loss by O-rich evolved stars, and so they allow us to evaluate 
our optically thicker O-rich models.  We applied zero-point offsets to 
2MASS data \citep[as recommended by][]{coh03}.  Note that we do not 
attempt to match our model grid for O-rich evolved stars to observed 
C-rich AGB stars and candidates.  A model grid for C-rich AGB stars is 
being computed in order to match observed data for C-rich AGB stars 
and candidates (Paper V).

We must caution that, though these comparisons between grid models 
and data in CMDs and CCDs show consistency between our model grid 
as a whole and observed data, one must be careful not to draw too many 
conclusions from these model/data comparisons in CMDs and CCDs.  
Though we have made an effort to avoid unrealistic models (e.g., no 
models with dust temperature at dust shell inner radius greater than dust 
sublimation temprtature, dust shell inner radius in the range expected by 
theory, etc.), certain combinations of model T$_{\rm eff}$, R$_{\rm min}$, 
$\tau_{\rm 10}$, and L$_{\rm star}$ may be unlikely or even unphysical.  
However, we wish to cover a somewhat larger range of the parameters 
varied in our grid than expected.

\subsection{CMD and CCD Comparisons of Model Grid to Data}

\subsubsection{Color-Magnitude Diagrams}

Color-magnitude diagrams are more useful in separating populations of sources 
with different luminosities.  We use this to our advantage here to draw the 
RSG candidate population away from the O-rich AGB candidate population and 
evaluate how our model grid covers each population, in addition to the extreme 
AGB candidate population.  Our model grid is consistent with almost all of the 
observed data in the CMDs for O-rich AGB star candidates from the SAGE sample, 
the O-rich AGB stars from SAGE-Spec, RSG star candidates from SAGE, RSG stars 
from SAGE-Spec, the OH/IR stars from \citet{marsh04}, and the extreme AGB 
candidates from SAGE.

\subsubsubsection{K$_{s}$ versus J-K$_{s}$}

The K$_{s}$ versus J-K$_{s}$ CMD is quite useful to discuss, as 
we use it in identification of O- and C-rich AGBs in the SAGE sample 
(see discussion in Section 2.1).  Figure 2 shows the objects identified 
as candidate O-rich AGB stars, C-rich AGB stars, RSGs, and extreme AGB 
stars in the SAGE sample plotted as blue, purple, red, and green points, 
respectively.  The O-rich AGB candidates form a 
nearly vertical column of slightly positive slope, with the density of 
points decreasing to lower K$_{s}$ magnitudes.  At K$_{s} \ltapp$ 9, 
the RSG candidates form a column approximately parallel to but offset from the 
O-rich AGB candidates.  At the same range of K$_{s}$ magnitudes but with 
slightly redder J-K$_{s}$ colors are the C-rich AGB stars.  Toward redder 
J-K$_{s}$ colors than the C-rich AGB candidates but at 
similar K$_{s}$ magnitudes to them are the extreme AGB candidates.  Also 
plotted are points corresponding to our model grid, as small black 
points.  We note that the black points line up roughly in columns, 
corresponding to sets of models with the same T$_{eff}$, R$_{min}$, and 
$\tau_{10}$ but different luminosity.

Our models reproduce the range of near-infrared colors and magnitudes 
observed in the SAGE and SAGE-Spec samples quite well.  All 17\,956 
O-rich AGB candidates identified from SAGE photometry as well as the 32 
AGB stars whose O-rich chemistries have been confirmed from SAGE-Spec 
spectra fall within the coverage of our models.  We have finer luminosity 
resolution in the range 10$^{3}$ to 2$\times$10$^{4} \lsun$ by design, 
corresponding to the greatest concentration of O-rich AGB candidate points 
(for K$_{s} \gtapp$ 9.5).  This is in anticipation of future efforts to model the 
SED of each O-rich AGB star candidate in detail, with the intention to constrain 
accurately the luminosity of each candidate O-rich AGB star.  Further, the points 
plotted as small gold stars, identified as O-rich AGB stars in the SAGE-Spec 
sample \citep{woods10}, are also well-matched by the model grid, with the 
exception of three points at (J-K$_{s}$, K$_{s}$) of (1.46, 9.36), (1.90, 10.41), 
and (2.04, 12.90).  These three points belong to MSX LMC 61, SSTISAGE1C 
J053128.42-701027.2, and SSTISAGE1C J045128.56-695550.1, respectively, 
all of which have SAGE-Spec spectra showing emission features at 10 and 
20$\mum$ wavelength belonging to silicates, confirming their O-rich natures.  
This mixing of O-rich AGB stars in this CMD in the region of moderately red 
colors dominated by C-rich AGB stars has been noted before \citep{marig08}.  
The brown triangles, representing the OH/IR stars from \citet{marsh04}, are 
also all covered by our model grid.  These 8 OH/IR stars lie at redder colors than 
the vast majority of the O-rich AGB stars and RSG stars, and, with the exception 
of IRAS 05280-6910, they have K$_{s} \ltapp$ 10.5.  IRAS 05280-6910, located 
at about (1.5, 12.8), is somewhat isolated from other O-rich stars but is near the 
SAGE-Spec O-rich AGB star SSTISAGE1C J045128.56-695550.1.  Two of the 
10 OH/IR stars from \citet{marsh04} have no {\it J} magnitudes in the SAGE 
catalog and are not plotted.  The magenta star symbols are C-rich AGB stars 
from the SAGE-Spec sample, which, like the C-rich AGB stars from SAGE, 
should not be expected to be covered by our O-rich grid.

The model grid covers 107 of the 108 RSG stars in the SAGE sample 
and all 18 RSGs in the SAGE-Spec sample.  The one SAGE RSG star not 
covered by the grid has a J-K$_{s}$ color only $\simali$0.01 magnitude less 
than the bluest models in the grid.  According to \citet{skrut06}, the 1-$\sigma$ 
uncertainty for this point, which lies near (J-K$_{s}$, K$_{s}$) = (0.6, 7.8), in 
K$_{s}$ is $\simali$0.01 magnitude for K$_{s}$ = 7.8, and the uncertainty in J is 
$\simali$0.015 magnitude for J = 8.4.  The 3-$\sigma$ uncertainty in the 
J-K$_{s}$ color is then $\simali$0.05, so this uncovered RSG star from SAGE is 
consistent with the model grid.  The RSG candidates in the 
SAGE-Spec sample, plotted as filled small orange circles, land on top of the 
densest concentration of SAGE RSG candidate points and are fully covered 
by the grid, further confirming the good coverage of the 
RSGs by our model grid.  Our grid also covers almost all of the extreme AGB 
star candidates (1123/1125).  Only two are outside of the range of our model 
grid, but we note that extending the grid to lower luminosities would allow 
these two points to be covered by our grid.  However, as the majority of 
extreme AGB stars shown in this CMD may not have O-rich dust chemistry 
(see later discussion), it may be of little meaning that lower luminosity models 
would cover these two points.  In the future, we hope to use SED-fitting as a 
means to distinguish O-rich from C-rich dust chemistry among the extreme 
AGB sample.

\subsubsubsection{[8.0] versus [3.6]-[8.0]}

Qualitatively, this one is similar to the K$_{s}$ versus J-K$_{s}$ CMD 
discussed previously.  The O-rich AGB candidates are faintest and bluest 
in color, while the RSG candidates have similar color but are brighter.  The 
C-rich AGB star candidates have brightnesses similar to the O-rich AGB 
candidates but have redder colors, while the extreme AGB star 
candidates are somewhat brighter and much redder.  We note the O-rich AGB 
candidate points now form a ``wedge'', the pointed end of which is at the 
bluest colors and highest magnitudes, and opposite end of which extends 
toward red colors and the lowest magnitudes.  There is a slight overlap 
between the brightest O-rich AGB candidate points in this ``wedge'' and the 
redder RSG candidates.  We note, however, moderate overlap between the O-rich 
AGB and extreme AGB candidates in the SAGE sample around [3.6]-[8.0] 
$\simali$0.5 and [8.0] $\simali$9.  Here, the extreme AGB candidates' [8.0] 
magnitudes appear to be similar to the brightest [8.0] magnitudes of the O-rich AGB 
candidate stars.  Again we note the overlap between RSG and O-rich AGB 
candidates for the reddest colors.  The isolated O-rich AGB star from SAGE-Spec 
at ([3.6]-[8.0],[8.0]) of (2.39, 5.08) is IRAS 05298-6957.  This star is also in the 
\citet{marsh04} OH/IR list and so is plotted here as a brown triangle.  It has silicate 
absorption features and is therefore confirmed to have O-rich dust.  The data 
point for this star on this CMD is in the same location as the region in the same 
CMD shown by \citet{matsu09} occupied by a handful of spectroscopically-confirmed 
O-rich AGB or RSG stars intruding into the region dominated by C-rich AGB stars.

Again, the model grid is very successful in matching the regions spanned by the 
O-rich AGB star candidates (from both SAGE and SAGE-Spec samples), RSG 
candidates (SAGE and SAGE-Spec), and extreme AGB candidates.  The O-rich 
AGB candidates from SAGE with the bluest 
[3.6]-[8.0] colors are missed by the model grid, as only 9507 out of 17783 of 
these are directly covered by the model grid.  However, of the 8276 that are not 
directly covered by the model grid, all but 2 of these are consistent within 3$\sigma$ 
errors with the model grid.  The 1$\sigma$ error for stars of [3.6] $\simali$12 in the 
IRAC 3.6$\mum$ band in SAGE Epoch 1 is $\simali$0.1 magnitude\footnote{see 
http://irsa.ipac.caltech.edu/data/SPITZER/SAGE/doc/SAGEDataDescription\_Delivery2.pdf}, 
and [3.6]-[8.0] $\sim$ 0 for these, so the 3$\sigma$ error in the [3.6]-[8.0] color would 
be $\simali$0.4.  There are about 6 O-rich AGB candidates in the SAGE sample with 
colors bluer than -0.3 (approximately the bluest color on this CMD that could be 
consistent with the model grid), but there are also 2 C-rich AGB candidates with 
similarly blue colors, so these 6 O-rich AGB candidates in SAGE may be misidentified.  
The model grid covers 32 out of 34 O-rich AGB stars from SAGE-Spec, 103 out of 109 
RSGs from the SAGE sample, and 16 out of 18 RSGs from the SAGE-Spec sample.  
The few RSGs from SAGE and SAGE-Spec not covered by the 
model grid have [8.0]$\simali$ 9 and zero [3.6]-[8.0] color, so they have [3.6]$\simali$ 
9 as well.  At these magnitudes, the 1$\sigma$ errors on the [3.6] and [8.0] magnitudes 
should be about 0.05 and 0.03 magnitudes, respectively, so the 3$\sigma$ error on the 
[3.6]-[8.0] color for the SAGE and SAGE-Spec RSGs could be up to 0.17 magnitude.  
Similarly, the two SAGE-Spec O-rich AGB stars have [8.0] $\sim$ [3.6] $\sim$ 10.  For 
these, the typical uncertainties on the [3.6]-[8.0] color would be slightly higher.  
The 2 O-rich AGB stars and 2 RSG stars from SAGE-Spec and the 6 RSGs from SAGE 
not directly covered by the grid are therefore still consistent with the model grid.

All 10 of the OH/IR stars from \citet{marsh04} are covered by the grid.  Most of these 
cluster around ([3.6]-[8.0],[8.0]) = (2, 6), while the one near (5,4.5) is IRAS 05280-6910, 
and the one near (2.4, 2.7) is IRAS 04553-6825 (WOH G64).  In the SAGE sample, 
1411 out of 1427 extreme AGB star candidates are covered by 
the model grid in this CMD.  Those that are not are redder than the reddest models in the grid.  
These extreme AGB candidates fall in the same region of this CMD as C-rich AGBs 
from the SAGE-Spec sample, which suggests these reddest extreme AGB star 
candidates on this CMD have carbon-rich dust.  Further, this suggests our grid of O-rich models 
should not be expected to match the reddest extreme AGB candidates on this plot.

\subsubsubsection{[24] versus [8.0]-[24]}

The [24] versus [8.0]-[24] CMD, plotted in Figure 4, differs from the rest of the 
CMDs we have discussed.  Here we see the O-rich AGB candidates branch into 
two arms at the reddest [8.0]-[24] colors, a brighter population and a fainter 
population.  The fainter population was identified as ``F'' in the same CMD shown 
by \citet{blum06}.  The RSG candidates extend from the brighter arm.  The extreme 
AGB candidates, however, now appear brighter than both the O-rich AGB candidates 
and the RSG candidates.  The mid-IR fluxes of xAGB stars are dominated by contribution 
from circumstellar dust, and their SEDs hence peak at longer wavelengths than those of 
optically thin O-rich AGB stars or RSG stars.  Again, the isolated O-rich AGB 
star from the SAGE-Spec sample at ([8.0]-[24],[24]) of (3.00, 2.09) is IRAS 05298-6957 
\citep[again, also from the OH/IR list of][]{marsh04}, confirmed to have O-rich dust, and 
it is also plotted here as a brown triangle.

We include in Figure 4 four tracks with points plotted on them.  Each track represents 
a set of models for which all parameters are the same except for $\tau_{\rm 10}$, which 
is allowed to vary.  The effect of varying $\tau_{\rm 10}$ is demonstrated within each 
track.  Each yellow circle on a track is for a different value of $\tau_{\rm 10}$, with the 
lowest values (optically thin models) being at the bluest colors and highest values 
(optically thick models) being at the reddest colors.  These tracks show that varying 
$\tau_{\rm 10}$ has a large effect upon the [8.0]-[24] color.  These tracks are not evolutionary 
models; instead, they demonstrate the effect of chaning the various parameters on the 
colors and magnitudes of the models.  \citet{ivez95} performed a similar exercise, with 
IRAS color-color diagrams instead of the CMD we explore here, to explore varying the 
optical depth of their dust shell models on model colors.  From bottom (greatest [24] 
magnitude) to top (lowest [24] magnitude), the tracks are for models with the following 
combinations of T$_{\rm eff}$, R$_{\rm min}$, and luminosity: 2100\,K, 15$\rstar$, and 
4820$\lsun$; 3700\,K, 15$\rstar$, and 16000$\lsun$; 2100\,K, 7$\rstar$, and 
57000$\lsun$; and finally 2700\,K, 7$\rstar$, and 124000$\lsun$.

T$_{\rm eff}$ has an effect like that of optical depth but lesser in strength.  Namely, 
cooler stars give rise to the reddest models 
for the optically thickest sources (bottommost track), while warmer stars give rise to the 
bluest models for the optically thinnest sources (the track that is second from bottom).  
Changing the luminosity translates a track vertically without distortion, as the flux at 
each wavelength in a model is scaled by the same scalar to generate a new model with 
all parameters the same except for the new luminosity (equal to the old luminosity 
multiplied by the aforementioned scalar).  This exact effect is not shown, but note that, 
from bottom to top, the tracks correspond to increasing values of luminosity.  The slope of each 
track changes around the range of [8.0]-[24] colors of 2--4.  This inflection is most likely 
due to the flattening of the model SED at mid-infrared wavelengths ($\simali$10$\mum$) 
due to the dust shell becoming optically thick at these wavelengths and also the 
10$\mum$ silicate feature transforming from being in emission for $\tau_{\rm 10} <$ 1 
to absorption when $\tau_{\rm 10} >$ 1.  The effect of varying R$_{\rm min}$ is 
demonstrated by the lowest and second-highest tracks.  For these two tracks, 
T$_{\rm eff}$ is the same, but the dust shell inner radius is 15$\rstar$ for the lower one 
and 7$\rstar$ for the higher one.  For dust shells of $\tau_{\rm 10}$ = 1 (indicated for a 
given track by a yellow circle with a red X overlaid), the dust shell with R$_{\rm min}$ = 
7$\rstar$ has hotter dust overall; therefore, it lies at a bluer color than the one with 
R$_{\rm min}$ = 15$\rstar$ and $\tau_{\rm 10}$ = 1.

The coverage of O-rich AGB candidates, RSG candidates, and extreme AGB 
candidates by the model grid for this CMD is similar in quality to that for the previous 
one, [8.0] versus [3.6]-[8.0].  Most of the three populations are covered by the model 
grid, except for stars with the bluest and reddest [8.0]-[24] colors and largest [24] 
magnitudes.  At bluer colors than the blue edge of the model grid coverage lie 186 
out of the 5564 O-rich AGB star candidates from SAGE, 3 out of 105 RSG candidates 
from SAGE, and 28 of the 1352 SAGE extreme AGB star candidates.  At [24]$\simali$ 
11, the 1$\sigma$ error in [24] is $>$0.2, probably up to 0.5 magnitudes; for [8.0]-[24] 
colors around -0.4, [8.0] is about 10.6, and the 1$\sigma$ error is about 0.05 (again, 
see the SAGE Data Delivery document).  The 3$\sigma$ error on the [8.0]-[24] 
color for the faint O-rich AGB SAGE candidates bluer than the model grid could 
therefore be up to 1.5 magnitudes, so the model grid is consistent with all these 
candidates.  For [24] = 8.5 and [8.0]-[24] = 0, the 3$\sigma$ error on [8.0]-[24] could be 
up to 0.2 magnitudes.  Thus, the model grid is consistent with all extreme AGB candidates 
except 1 and all SAGE RSG stars.  Note that this does not mean 
the extreme AGB candidates predominantly have O-rich dust.  In fact, our later discussion 
suggests the opposite, that the extreme AGB stars predominantly have C-rich chemistry, 
including the extreme AGB candidate that is too blue in color to be consistent with our 
O-rich grid (see discussion in previous subsection 
on the [8.0] versus [3.6]-[8.0] CMD regarding the bluest observed points).  Beyond the 
reddest models, there are 9 extreme AGB candidates and 1 O-rich AGB candidate from 
SAGE.  The O-rich AGB candidate suggests that some of the 9 extreme AGB candidates 
may have oxygen-rich dust chemistry (in which case $\tau_{\rm 10}$ would be greater than 
26), but the presence of 2 C-rich AGB stars from the 
SAGE-Spec sample in this region of the CMD and the presence of C-rich AGB stars 
from SAGE-Spec in the same region of the previous CMD suggests the dominant 
dust chemistry for extreme AGB candidates of these red colors to be carbon-rich.  All 
34 O-rich AGB stars from SAGE-Spec and all 17 RSG stars from SAGE-Spec are 
within the model grid coverage.  The 8 OH/IR stars from \citet{marsh04} are all covered 
by the grid and are located near (3, 3).  A handful of about 10 O-rich AGB star candidates 
from the SAGE sample with 2 $<$ [8.0]-[24] $<$ 4 and 6 $<$ [24] $<$ 10 lie at fainter 
[24] than is covered by the model grid.  These likely belong to a third population of 
either red giant branch (RGB) stars fainter than the tip of the RGB or O-rich AGB stars 
that have sufficiently low mass that they will not become C-rich AGB stars in the future.  
Were our model to extend to luminosities lower than 10$^{3} \lsun$, it could cover these.

\subsubsection{Color-Color Diagrams}

Color-color diagrams, also known as two-color diagrams, are useful in identifying 
populations of sources with distinctive changes in broadband colors over a wide 
wavelength range in an SED.  The two colors trace this change in color with 
wavelength.  Again, our model grid is consistent with almost all of the observed 
data for O-rich AGB star candidates from the SAGE sample, the O-rich AGB stars 
from SAGE-Spec, RSG star candidates from SAGE, the RSG stars from SAGE-Spec, 
and the OH/IR stars from \citet{marsh04}.  The extreme AGB candidates from SAGE, 
however, are frequently not consistent with our model grid in the CCDs, which has 
implications for the dust chemistry of the extreme AGB candidates.

\subsubsubsection{[5.8]-[8.0] versus [3.6]-[4.5]}

This color-color diagram, plotted in Figure 5, traces the change in color over 
the near-infrared wavelength range of 3--8$\mum$ spanned by the {\it Spitzer} 
IRAC bands.  The most obvious feature of this CCD is that, in general, the redder 
a source is in one color, the redder it is in the other.  This is due to both colors 
measuring the same general component - the emission from the circumstellar 
dust shell.  The next most obvious characteristic of this plot is that those sources 
identified as extreme AGB candidates are generally redder in both colors than 
both the O-rich AGB and RSG candidates.  This suggests that the extreme AGB 
candidates generally have more dust in their dust 
shells than either the O-rich AGB or RSG candidates.  Additionally, the O-rich 
AGB and RSG candidates overlap greatly in the plot.  This suggests the shell 
properties or dust properties (or both) are similar between the two populations, 
and that the more significant difference between the two is their luminosities 
(see previous discussion on RSG and O-rich AGB candidates in the CMDs).

Here, the coverage of the observed data by the model grid is quite interesting.  
At the bluest colors, the model grid begins in the middle of the densest 
concentration of O-rich AGB candidate points at the bluest colors and also 
close to the bluest RSG candidates.  The model grid then extends in the 
direction of the reddest O-rich AGB candidates (as the O-rich AGB stars in 
SAGE-Spec and the RSGs from both the SAGE and SAGE-Spec samples do) 
until it reaches [3.6]-[4.5]$\simali$ 0.3 and [5.8]-[8.0]$\simali$ 1.3.  At this point, 
the model grid abruptly turns to redder [3.6]-[4.5] colors, with [5.8]-[8.0] color 
not changing much.  It is also seen that the C-rich AGB stars in the SAGE and 
SAGE-Spec samples are located in regions occupied by the densest 
concentrations of extreme AGB candidate points.  One region is the ``finger'' of 
extreme AGB points extending from point ([3.6]-[4.5], [5.8]-[8.0]) = (0.3, 0.3) to 
(-0.2, 0.7), which is almost perpendicular to the finger of RSG candidate points.  
The other, more densely populated concentration of extreme AGB candidate 
points, in which are found many C-rich AGBs from the SAGE-Spec sample, 
extends from (0.3, 0.3) to (1, 0.7).  This suggests a large majority of sources 
in the SAGE sample identified as extreme AGBs have carbon-rich chemistry.  
As with the CMDs, there is an exception in the isolated O-rich AGB star IRAS 
05298-6957 from the SAGE-Spec sample \citep[and OH/IR sample;][]{marsh04} 
at ([3.6]-[4.5],[5.8]-[8.0]) of (0.80, 0.91).

Amongst the O-rich AGB stars in the SAGE-Spec sample, the model grid 
provides mostly good coverage except for the bluest such sources.  From the 
CMDs, we learn that the bluest sources are often the faintest, so we apply 
those lessons here.  In the [8.0] versus [3.6]-[8.0] CMD, the faintest stars had 
[3.6]$\simali$ 12 and [8.0]$\simali$ 12.3.  Assuming [3.6]-[4.5] = [5.8]-[8.0] = 0, 
then [4.5] = 12 and [5.8] = 12.3 typically for the blue O-rich AGB SAGE 
candidates not covered by the grid in this CCD.  According to the SAGE Data 
Delivery document, the 1$\sigma$ errors in [4.5] and [5.8] for these bluest stars 
are 0.04 and 0.07 magnitudes, respectively.  This means the 3$\sigma$ 
error in [3.6]-[4.5] could be up to $\simali$0.3 and that in [5.8]-[8.0] could be 
up to $\simali$0.4.  Though strictly speaking only 3436 out of the 17769 O-rich 
AGB candidates from SAGE plotted here are covered by the model grid, all of 
the 17769 except the 6 with [3.6]-[4.5] $<$ -0.4, the $\simali$11 with [5.8]-[8.0] 
$<$ -0.35, and the $\simali$30 with [3.6]-[4.5] $>$ 0.4 separated from the 
densest concentration of O-rich AGB candidates from SAGE and bluer than 
the model grid are consistent with the model grid.  The 17 bluest stars (6+11) 
not consistent with the model grid may be misidentified and may actually be 
C-rich AGB stars (see discussion in CMD subsections about the stars with 
the bluest colors).  Similarly, the group of 30 may also be misidentified, as it 
has colors more like C-rich AGB stars from SAGE-Spec, C-rich AGB 
candidates from SAGE, and extreme AGB candidates (which we have 
already noted are more likely to have carbon-rich dust than oxygen-rich dust) 
from SAGE.

Only 38 out of 107 RSGs from SAGE, 21 out of 32 O-rich AGB candidates from 
SAGE-Spec, and 4 out of 18 RSGs from SAGE-Spec are covered by the grid.  
However, all the uncovered ones from these three populations are at most 
only 0.3 magnitudes separated from any edge of the grid coverage, so 
they are consistent with the model grid.  Seven of the 10 OH/IR stars are 
covered by the grid.  Two of the \citet{marsh04} OH/IR stars lie very close to the 
grid but just to slightly bluer [5.8]-[8.0] colors of it, IRAS 04545-7000 and IRAS 
05003-6712.  The edge of the grid is consistent with these two OH/IR stars, lying 
within their 3-$\sigma$ errors.  However, IRAS 05280-6910 is neither covered 
by the grid nor consistent with it, being located near (1.5, 2).  The reason for 
this discrepancy is unknown, but we do note that the {\bf 2D}ust model of this 
star's SED by \citet{boyer10} indicates a very large total mass-loss rate from this 
star of a few times 10$^{-3} \msunyr$, consistent with this OH/IR star's very red 
colors in this CCD.  Lastly, only 307 out of the 1423 extreme AGB candidates 
from SAGE plotted are covered by the model grid.  However, as we noted 
previously, the extreme AGB stars may be more consistent with a carbon-rich 
dust chemistry than oxygen-rich dust chemistry.

\subsubsubsection{K$_{s}$-[8.0] versus K$_{s}$-[3.6]}

In a very general sense, the K$_{s}$-[8.0] versus K$_{s}$-[3.6] CCD, plotted in 
Figure 6, is similar to the previous one, [5.8]-[8.0] versus [3.6]-[4.5].  At the bluest 
colors lie the O-rich AGB candidates.  On top of them lie the RSG candidates from 
SAGE as well as both the O-rich AGB stars and RSGs from SAGE-Spec.  These, 
as previously, form a locus of points around a line of highly positive slope.  At 
redder colors lie the extreme AGB candidates.  The O-rich AGB star from the 
SAGE-Spec sample \citep[and OH/IR sample;][]{marsh04}, IRAS 05298-6957, 
lies at (3.87, 6.26), in the midst of the region dominated by extreme AGB candidate 
stars.

The bluest models provide good coverage of the RSG candidates from SAGE, 
the RSGs from SAGE-Spec, the densest concentration of O-rich AGB candidates 
from SAGE, and the O-rich AGB stars from SAGE-Spec.  The coverage is better 
than for the [5.8]-[8.0] versus [3.6]-[4.5] CCD, though a number of O-rich AGB, 
extreme AGB, and RSG candidates from SAGE are not covered by the model 
grid.  Using 1$\sigma$ errors for the faintest stars cited in previous subsections 
for [3.6] and [8.0] and assuming a 1$\sigma$ uncertainty of 0.02 magnitude for 
K$_{s}$=12 \citep{skrut06}, the 3-$\sigma$ error for both K$_{s}$-[8.0] and 
K$_{s}$-[3.6] colors could be up to 0.3 magnitudes.  Thus, all 17783 O-rich 
AGB candidates from SAGE plotted except $\simali$ 30 with K$_{s}$-[3.6] $<$ 
-0.2 and another $\simali$ 20 with K$_{s}$-[3.6] $\sim$ 1.1 and K$_{s}$-[8.0] 
$\sim$ 1.5 are consistent with the model grid.  As concluded for the CMDs and 
for the previous CCD, the $\simali$ 30 with the bluest colors may be 
misidentified.  The $\simali$ 20 with intermediate colors lie directly on top of a 
dense population of C-rich AGB candidates from SAGE (purple points), so they 
also may be misidentified as O-rich.  Of the 108 RSG candidates from SAGE 
plotted, all but 1 are consistent with the grid.  The 1 inconsistent one, with 
K$_{s}$-[3.6]$\simali$ -0.25, has a very blue color and may be misidentified.  The 
33 O-rich AGB stars and 18 RSG stars from SAGE-Spec plotted are all consistent 
with the model grid.  In this CCD, 9 out of the 10 OH/IR stars from \citet{marsh04} 
are covered by the grid, and the discrepant one again is IRAS 05280-6910, 
which, as we noted earlier, has a very high mass-loss rate \citep{boyer10}.

As with the previous CCD, a majority of the extreme AGB candidates are not 
covered by the model grid.  The model grid covers two O-rich AGB stars from 
SAGE-Spec at K$_{s}$-[3.6] colors of about 2 and 4, which are located in the middle 
of the extreme AGB region on the plot.  These two O-rich AGB stars provide a 
test of the model grid at high optical depth, which it passes quite successfully.  
The model grid suggests a lower envelope to the K$_{s}$-[8.0] color of any extreme 
AGB stars with O-rich dust chemistry.  This lower envelope can be described 
by the lines

\begin{eqnarray}
K_{s}-[8.0]~~= & 5.13\times(K_{s}-[3.6]) - 2.38 &~~~~{\rm for}~~~~~~K_{s}-[3.6] < 0.95 ~~ {\rm and} \nonumber \\
K_{s}-[8.0]~~= & 1.28\times(K_{s}-[3.6]) - 1.22 &~~~~{\rm for}~~~~~~K_{s}-[3.6] \geq 0.95 ~,
\end{eqnarray}

\noindent such that the extreme AGB stars with lower K$_{s}$-[8.0] colors than these 
lines probably do not have O-rich dust chemistry, having C-rich dust chemistry 
instead.  In total, 525 out of the 1283 extreme 
AGB candidates from SAGE plotted have K$_{s}$-[8.0] colors higher than this envelope.  
As C-rich AGB stars from SAGE-Spec are above the envelope in the extreme AGB 
region of the CCD, 525 is most likely an upper limit on the number of extreme AGB 
candidates in this plot that could have O-rich dust chemistry.  Finally, we note the 
O-rich and C-rich AGB stars in the SAGE-Spec sample 
separate fairly well for K$_{s}$-[3.6] $\ltapp$ 1.3; for redder K$_{s}$-[3.6] colors, AGB 
stars of the two chemistries in the SAGE-Spec sample mix together on the 
plot.  Over most of the K$_{s}$-[3.6] colors spanned by the data plotted in Figure 6, 
the O-rich models have redder K$_{s}$-[8.0] colors than most of the extreme AGB candidates 
from SAGE, which overlap with a number of C-rich AGB stars from the SAGE-Spec 
sample.  The O-rich stars having redder K$_{s}$-[8.0] colors is most likely due to the O-rich 
AGB stars having lower continuum emission over near-infrared wavelengths ($<$ 
8$\mum$) relative to the flux from IRAC 8$\mum$ \citep[see discussion by][]{sarg10} 
than is true for C-rich AGB stars (for which the dust emissivity monotonically 
decreases; Srinivasan et al., {\it in prep}).  However, we note that, in general, this 
plot is less useful in separating the O-rich and C-rich evolved star populations in 
the SAGE-Spec sample than the [5.8]-[8.0] versus [3.6]-[4.5] CCD was.

\subsubsubsection{[3.6]-[24] versus K$_{s}$-[3.6]}

The final CCD we discuss, that of [3.6]-[24] versus K$_{s}$-[3.6] plotted in Figure 
7, is somewhat different from the other two CCDs discussed here.  The 
lower envelope to the [3.6]-[24] colors of the reddest models is described 
by the lines

\begin{eqnarray}
[3.6]-[24]~~= & 7.5\times(K_{s}-[3.6]) - 3.8 &~~~~{\rm for}~~~~~~K_{s}-[3.6] < 0.99 ~~ {\rm and} \nonumber \\\relax
[3.6]-[24]~~= & 0.55\times(K_{s}-[3.6]) + 3 &~~~~{\rm for}~~~~~~K_{s}-[3.6] \geq 0.99 ~,
\end{eqnarray}

\noindent where only 147 out of the 1223 extreme AGB candidates from 
SAGE plotted are above the lower envelope.  Here, the O-rich AGB 
candidates stretch out in the direction of [3.6]-[24] while only narrowly 
spreading out over a very small range of K$_{s}$-[3.6].  As before, the 3-$\sigma$ 
error on K$_{s}$-[3.6] is $\simali$0.3.  The 1-$\sigma$ error on [24] for the 
faintest sources is up to $\simali$0.5, while that for [3.6] is $\simali$0.1 (see 
previous discussion), so the 3-$\sigma$ error on [3.6]-[24] is up to $\simali$1.5.  
The 5566 O-rich AGB candidates from SAGE plotted are consistent with the model 
grid except for three groups.  The $\simali$ 10 O-rich AGB SAGE candidates at the 
bluest K$_{s}$-[3.6] colors are likely misidentified, the $\simali$3 at the reddest 
[3.6]-[24] colors may suggest different T$_{\rm eff}$ or R$_{\rm min}$ than used in 
the model grid, and the $\simali$ 30 with K$_{s}$-[3.6] $\sim$ 1 and [3.6]-[24] 
$\sim$ 2.5 underneath the lower envelope may also be misidentified.

As for the [24] versus [8.0]-[24] CMD, we plot tracks for sets of models 
having all parameters at the same value except for $\tau_{\rm 10}$, and we indicate 
the point for which $\tau_{\rm 10}$ = 1.  As with the [24] versus [8.0]-[24] CMD, the 
tracks show that $\tau_{\rm 10}$ has the greatest effect on the models, causing 
their colors to span almost the full range of both K$_{s}$-[3.6] and [3.6]-[24] colors 
covered by our model grid.  As with the tracks on the CMD, the track bends at 
intermediate colors (K$_{s}$-[3.6] of $\simali$0.5) going from nearly vertical 
to a low positive slope.  This bend is most likely due to the flattening of the model 
SED at mid-infrared wavelengths ($\simali$10$\mum$) due to the dust shell 
becoming optically thick at these wavelengths.  For the track that begins furthest left 
(lowest K$_{s}$-[3.6] color) and ends with the lowest [3.6]-[24] colors, the parameters 
are set at T$_{\rm eff}$ = 3700\,K, R$_{\rm min}$ = 15$\rstar$, and luminosity of 
16000$\lsun$.  The two tracks that begin at right (K$_{s}$-[3.6] $>$ 0.5) both share 
T$_{\rm eff}$ = 2100\,K, which shows, as the [24] versus [8.0]-[24] CMD showed, that 
T$_{\rm eff}$ has only a secondary effect on color, mildly altering the K$_{s}$-[3.6] 
color for the optically thinnest models (the points with the lowest [3.6]-[24] colors).  
Of the two tracks that begin at right, the one that reaches the greatest [3.6]-[24] colors 
for the optically thickest models (furthest to the right, or greatest K$_{s}$-[3.6] colors) 
has R$_{\rm min}$ = 15$\rstar$ and luminosity of 4820$\lsun$.  The track that begins 
at right but ends at lower [3.6]-[24] colors for the optically thickest models has 
R$_{\rm min}$ = 7$\rstar$ and luminosity of 57000$\lsun$.  These last two tracks 
demonstrate the effect of R$_{\rm min}$ on this CCD.  The [3.6]-[24] color of the 
model with $\tau_{\rm 10}$ = 1 and the greater R$_{\rm min}$ has a greater 
[3.6]-[24] color than the model with $\tau_{\rm 10}$ = 1 and the lesser R$_{\rm min}$ 
because the model with greater R$_{\rm min}$ has cooler dust on average.  Again, 
varying luminosity but no other parameters has no effect on CCDs because all 
wavelengths are multiplied by the same scalar to transform from a model with 
one luminosity to another.

We determined that the [3.6]-[24] colors of bare photospheres (stars without 
dust shells) were only $\simali$0.3 magnitudes lower than models having 
shells with $\tau_{\rm 10}$ = 8$\times$10$^{-4}$.  The accuracy and precision 
with which we will be able to determine the mass-loss rates for the AGB 
candidates in the SAGE sample with the smallest excesses over stellar 
photosphere emission will therefore be limited.  However, we retain the models 
with $\tau_{\rm 10}$ between 1$\times$10$^{-4}$ and 8$\times$10$^{-4}$ 
because we will be better able to constrain mass-loss rates and other important 
parameters for stars with data having relatively small uncertainties.  
In subsequent studies we will use $\chi^{2}$-minimization to 
determine the best fitting model to each SAGE candidate AGB star, and this 
procedure will determine the best-fit values and uncertainties of various 
parameters of concern, including mass-loss rates.  Including these low optical 
depth models can only add to the value of the $\chi^{2}$-minimization analysis.

The RSG candidates form a nearly vertical column on top of the finger of 
O-rich AGB candidates extending from (0.3, 0) to (0.5, 3).  Again, all 104 
RSG candidates from SAGE plotted except the one with K$_{s}$-[3.6] of -0.25 
are consistent with the model grid, while the one with negative color may be 
misidentified.  The 17 RSGs and the 33 O-rich AGB stars from SAGE-Spec 
are consistent with the model grid.  In this CCD, all 8 of the OH/IR stars from 
\citet{marsh04} with sufficient photometry to be plotted here are covered by 
the model grid.

Interestingly, many of the O-rich AGBs in the SAGE-Spec sample with 
K$_{s}$-[3.6] $<$ 0.5 lie directly on top of the RSG candidate points.  Even more 
promising, a dense concentration of model points lands on top of these O-rich 
AGB SAGE-Spec points and RSG candidate points.  We interpret the stretching 
in the [3.6]-[24] direction of the O-rich AGB candidates, the RSG candidates, the 
bluest SAGE-Spec O-rich stars, and the bluest models in the grid as being due to 
the 3.6$\mum$ band being optimally placed between the shorter wavelengths 
where the stellar emission dominates, and the longer wavelengths where 
dust emission dominates.  Encouragingly, as with the other CCDs 
we discuss, the model points with the reddest K$_{s}$-[3.6] colors of 0.5 to 4 
match the SAGE-Spec O-rich AGB star points quite well.

\subsection{Verification of Model Grid by SED-Fitting}

As further verification of the model grid presented here, we fit models from 
our O-rich evolved star grid to the photometry of various stars, anticipating 
future work fitting our model grid to the thousands of candidate O-rich AGB, 
RSG, and extreme AGB stars in the SAGE sample.

\subsubsection{Optimal Model Fits to Selected Stars}

First, we fit the two O-rich 
AGBs modeled previously, HV 5715 and SSTSAGE052206 \citep{sarg10}.  
This provides a check on the modeling of those two stars.  To find the best fit 
of model to data from our model grid, we find the model from our final grid 
that has minimum $\chi^{2}$ with respect to the observed broadband 
{\it IJHK$_{\rm s}$}, IRAC, and MIPS 24$\mum$ photometry.  To allow for 
real stars' luminosities not being one of the 56 discrete values of luminosity 
in our final grid, we determine by another $\chi^{2}$ minimization the scalar 
that we multiply by our best-fit model flux to obtain the best possible fit.  This 
scalar is typically only a few percent away from unity.  This has the virtue of fitting models 
to photometric bands that have been used (except {\it I} and {\it H}) in CMDs 
and CCDs to verify the validity of the model grid presented here.  Here, 
$\chi^{2}$ is weighted by the photometric uncertainties.

For SSTISAGE1C J052206.92-715017.6 (``SSTSAGE052206''), the best fit 
obtained from our grid was with the model with luminosity L = 4900$\lsun$ 
and $\tau_{\rm 10}$ = 0.10.  The dust mass-loss rate for this model is 
2.1$\times$10$^{-9} \msunyr$.  This model is plotted in Figure 8 with the 
MCPS {\it I}, 2MASS {\it JHK$_{\rm s}$}, and epoch 1 catalog IRAC 
and MIPS-24 photometry plotted for this object by \citet{sarg10}.  Overall, 
these parameter values are not very far away from those determined by 
\citet{sarg10} when modeling multi-epoch broadband photometry and the 
{\it Spitzer}-IRS SAGE-Spec spectrum.  From that analysis, the best-fit model 
obtained had luminosity L = 5100$\lsun$ and $\tau_{\rm 10}$ = 0.095, with a 
dust mass-loss rate of 2.0$\times$10$^{-9} \msunyr$, so the dust mass-loss 
rate found here is only $\simali$5\% above the value obtain by our best fit 
here.  \citet{sarg10} were modeling additional data at {\it J}, 
{\it H}, and $K_{s}$ bands that closely agreed with the 2MASS data over 
these bands, which effectively weighted the importance of these bands more 
strongly in achieving a good fit.  Here, we only have the 2MASS data, so the 
near-infrared data overall constrains the best-fit model more weakly than 
that obtained by \citet{sarg10}, resulting in a lower luminosity in the best-fit 
model here.  In view of future studies, the most important parameters to 
determine are mass-loss rates and stellar luminosity.  For SSTSAGE052206, 
both dust mass-loss rate and luminosity are determined by fitting models from 
the O-rich model grid to match within $\simali$10\% of the values determined 
by \citet{sarg10}.

For HV 5715 the best fit obtained from our grid was with the model with 
stellar photosphere luminosity L = 33000$\lsun$ and $\tau_{\rm 10}$ = 
0.026.  The dust mass-loss rate for this model is 1.5$\times$10$^{-9} 
\msunyr$.  This fit is also shown in Figure 8.  As with the previous figure, the 
model is plotted with the MCPS {\it I}, 2MASS {\it JHK$_{\rm s}$}, and epoch 
1 catalog IRAC and MIPS-24 photometry plotted for this object by \citet{sarg10}.  
The match between the best-fit model from our model grid and the photometry 
is of similar quality to that for SSTSAGE052206.  The model presented for this 
object by \citet{sarg10} had luminosity L = 36\,000$\lsun$ and $\tau_{\rm 10}$ 
= 0.012.  The dust mass-loss rate found for this star by \citet{sarg10} is 
2.3$\times$10$^{-9} \msunyr$, so the dust mass-loss rate found here is 
$\simali$35\% lower.  This is most likely due to the restricted range of R$_{\rm min}$ 
that we chose for our model grid because it covers most of the range of dust 
shell inner radii expected for typical O-rich AGB stars \citep{hoef07}.  As noted by 
\citet{sarg10}, this star is quite variable, and if only the lowest fluxes were 
fit, the luminosity of the best-fit model would be about 31000$\lsun$, which 
is below what we get here.  This is understandable, as the 2MASS fluxes over 
{\it J}, {\it H}, and $K_{\rm s}$ that we fit here are some of the lower fluxes 
fit by \citet{sarg10}.  The relative error bars over the IRAC bands are larger 
than those for {\it J}, {\it H}, and $K_{\rm s}$ or 24$\mum$, which may 
account for the slight mismatching of model to data at 5.8 and 8.0$\mum$.  
The difference between the best fit to this object's data presented by 
\citet{sarg10} and that presented here may be understood by the additional 
photometry (and IRS spectrum) considered by \citet{sarg10}, which more strongly 
weights the near-infrared and mid-infrared parts of the SED than is done here.

We also explore the ranges of parameters explored by our model grid by fitting 
the SEDs of individual stars.  Figure 9 shows the best fits from our O-rich grid to 
the O-rich AGB star SSTISAGE1A J045947.31-694756.4 and the extreme AGB 
star SSTISAGE1A J053238.56-682522.2.  The O-rich AGB star has an extremely 
low dust mass-loss rate of 7.2$\times$10$^{-13} \msunyr$, while the extreme AGB 
star has a very high dust mass-loss rate of 6.7$\times$10$^{-7} \msunyr$.  This 
spans much of the range of dust mass-loss rates covered by our grid, as shown in 
Figure 1.  Dust mass-loss rate is directly influenced by $\tau_{\rm 10}$, so this 
justifies the large range of $\tau_{\rm 10}$ explored by our grid.

In Figure 10, we show best fits to two more stars - one with low luminosity and 
one with high luminosity.  The low luminosity star, SSTISAGE1A J052039.50-700344.3, 
is an O-rich AGB star, while the high luminosity star, Parker 1684, is a RSG star.  
The best-fit model to the former has a luminosity of 2260$\lsun$, while the best-fit 
model to the latter has a luminosity of 131000$\lsun$.  This justifies the large 
range of luminosities explored by the grid (Figure 1).

Figure 11 demonstrates the need for the range of stellar effective temperatures 
we explore in our grid.  SSTISAGEMC J053320.85-673031.6, a RSG star, is best fit 
by a model with T$_{\rm eff}$ = 4100\,K, while SSTISAGE1A J054041.69-661446.7, 
an O-rich AGB star, is best fit by a model with T$_{\rm eff}$ = 2500\,K.  In this Figure, 
one notes the obvious shift in the peak of the stellar emission from one star to 
the other.  This helps justify the range of T$_{\rm eff}$ explored in this grid.

Finally, in Figure 12 we demonstrate with two O-rich AGB stars the need for the 
wide range of R$_{\rm min}$ we explore in our grid.  SSTISAGE1A J052019.36-693529.0 
is best fit by a model with R$_{\rm min}$ = 3 $\rstar$, while SSTISAGE1A 
J052544.65-692740.4 is best fit by a model with R$_{\rm min}$ = 15 $\rstar$.  The 
relative contribution to the SED from warmer dust is greater than that for the latter.  
This explains why there is relatively more emission in excess of that from the stellar 
photosphere over the 4.5, 5.8 and 8.0$\mum$ bands with respect to the excess at 
24$\mum$ for the former star than for the latter star.  In fact, the observed photometry 
of the latter star suggests R$_{\rm min}$ perhaps somewhat larger than 15 $\rstar$, as 
the [8.0]-[24] color of the data is redder than that in the best-fit model.  However, we 
do not explore larger R$_{\rm min}$ due to theoretical considerations \citep{hoef07}.

\subsubsection{Comparison to Other Modeling Efforts}

In addition to justifying the range of parameters we explore in our model grid, we also 
test two of the most important parameters obtained from fitting SEDs with our models, 
luminosity and dust mass-loss rate.  We test the values we obtain for these against the 
values obtained for the same stars by \citet{vl99} and \citet{groen09} (only some of the 
stars in the two M-star target sets are common to both).  The coordinates for the M stars 
with and without {\it ISO} spectra from \citet{vl99} were matched against the SAGE 
epoch 1 catalog.  In all cases except two, the 3.6, 8.0, and 24$\mum$ fluxes of the 
closest SAGE source to the \citet{vl99} position matched fairly well the 3.6, 8.0, and 
24$\mum$ model fluxes plotted by \citet{vl99}.  The two for which the closest SAGE 
match did not match in these infrared fluxes were IRAS 05329-6708 and SP77 30-6.  
The former was much better matched to a SAGE source $\simali$27$\arcsec$ away 
than the one $<$1$\arcsec$ away.  The mid-infrared data fitted by \citet{vl99} for this 
source are a spectrum and photometry from the {\it Infrared Space Observatory} 
\citep[ISO;][]{kessler96}.  ISO had lower resolution than {\it Spitzer}, which may allow this 
large discrepancy in position.  The nearest SAGE match to the position of the latter 
target, SP77 30-6, was $\simali$8$\arcsec$ away, but the nearest SAGE source of 
comparable mid-infrared fluxes to those plotted by \citet{vl99} for this star was 
$\simali$14$\arcsec$ away.  We used the latter SAGE match as the data for this star.  
\citet{groen09} fit models to data that includes SAGE {\it Spitzer} data, so the matching 
between each \citet{groen09} position and the closest SAGE position is very likely the 
correct one.  We remove IRAS 04509-6922 from our analysis, as the only flux available 
for its match in the SAGE catalog that we consider is the 24$\mum$ flux, which is not 
enough to constrain its model.

In Figure 13 we compare luminosities obtained by the modeling done by \citet{vl99} and 
\citet{groen09} to those from the best fits of our model O-rich grid to each of the ones 
from \citet{vl99} and \citet{groen09} considered.  As may be noted in the plot, the 
luminosities obtained by {\bf 2D}ust modeling are within a factor of $\simali$2 of those 
obtained by \citet{vl99} or \citet{groen09}.  This is a fairly good correspondence, 
considering the factors that may contribute to obtaining different luminosities from our 
modeing versus the modeling by the other two studies.  These factors include variability, 
the use of different types of data \citep[we only use one epoch of photometry at each band in 
each SED, while the other authors also use spectroscopy;][additionally use photometry 
from multiple epochs]{groen09}, etc.

In Figure 14 we compare the dust mass-loss rates obtained by the two other modeling 
studies to those from our model grid.  The output of {\bf 2D}ust is dust mass-loss rate, 
so we convert the total mass-loss rates reported by \citet{vl99} and from \citet{groen09} to 
dust mass-loss rates by dividing their total mass-loss rates by their assumed gas-to-dust 
ratios of 500 and 200, respectively.  Here we see a difference between the target sets 
obtained from \citet{vl99} and from \citet{groen09}.  The matching between dust mass-loss 
rates obtained from our {\bf 2D}ust grid and those stars fit by \citet{vl99} is reasonable 
(a factor of $\simali$3) over many orders of magnitude of dust mass-loss rate.  As noted 
previously, the best SAGE match to the fluxes plotted by \citet{vl99} for IRAS 05329-6708 
come not from the closest SAGE match but instead from $\simali$27$\arcsec$ away 
(again, not implausible due to the lower ISO resolution).  Overall, our modeling agrees 
fairly well with that done by \citet{vl99}.

Figure 14 shows the dust mass-loss rates obtained by {\bf 2D}ust to be comparable to 
those obtained by \citet{groen09} at the highest dust mass-loss rates of $\simali$10$^{-7} 
\msunyr$.  However, at lower dust mass-loss rates the values obtained by {\bf 2D}ust 
begin to exceed those obtained by \citet{groen09}, building up to a factor of $\simali$6 
deviation for \citet{groen09} dust mass-loss rates near $\simali$2$\times$10$^{-9} 
\msunyr$.  It is difficult to identify the reason for this discrepancy.  We found that the 
optical depths at 10$\mum$ we obtain when modeling the \citet{groen09} targets are 
also greater than those obtained by \citet{groen09} for the ones with lower optical depths.  
For the ones with higher optical depths, the optical depths we obtain are similar to those 
\citet{groen09} obtained.  The pattern of disagreement between our modeling and that of 
\citet{groen09} is similar for both optical depth and dust mass-loss rate.  Both our modeling and 
that of \citet{vl99} do not fit data shortward of {\it I}-band, while the \citet{groen09} modeling 
does.  As an experiment, we tried fitting the {\it UBV} data in addition to the infrared data 
we already include in our fitting.  This resulted in slightly lower dust mass-loss rates, but 
not enough to explain the factor of 6 discrepancy.  Another difference is that \citet{vl99} 
and we only use a single type of dust in our modeling, while \citet{groen09} use multiple 
sources for their dust optical properties.  However, there was no apparent pattern showing 
different types of dust used by \citet{groen09} to correspond to any range of dust mass-loss 
rate.  \citet{groen09} used {\it Spitzer}-IRS spectra to constrain their modeling, while neither 
\citet{vl99} nor we did \citep[though][used ISO spectra]{vl99}.  The 8$\mum$ flux we use 
only partially covers the 10$\mum$ silicate band, so our only fitting photometry likely 
leaves our modeling less accurate when fitting low mass-loss rate stars' SEDs.  However, 
we note the few \citet{vl99} sources with low dust mass-loss rates agree fairly well with 
the dust mass-loss rates from our {\bf 2D}ust modeling.  When {\it Akari} \citep{mura07} 
data is made public, the addition of that data to the SAGE data will significantly help 
constrain our modeling of O-rich evolved stars.  We do note that \citet{srin09} found the 
highest mass-loss rate AGB stars contribute the most to the dust mass return to the LMC, so 
it is encouraging that our modeling agrees with that of both \citet{vl99} and \citet{groen09} 
for the highest dust mass-loss rate stars.  Finally, in Figure 15 we show the SED fits from 
our modeling to stars that were fit by both \citet{vl99} and \citet{groen09}.  These two stars 
are HV 888 and IRAS 05402-6956.  Our models of each of these fit the data quite well, 
which lends more support to the validity of our modeling.

\section{Conclusions}

With the goal of quantifying the mass-loss of O-rich evolved stars in the 
LMC in mind, we have constructed a grid of models of these stars.  To begin, 
we construct a grid of 1225 models using {\bf 2D}ust.  This grid 
assumes for each model a spherically symmetric shell of dust surrounding 
a star whose spectrum is assumed to be that of a PHOENIX model 
\citep{kucin05,kucin06} with stellar effective temperature between 2100\,K and 
4700\,K and log(g) = -0.5.  This initial grid of $\simali$1200 models has already 
excluded models whose dust temperature at the dust shell inner radius was above 
1400\,K on the basis of being unphysical, as this is approximately the sublimation 
temperature of silicate dust.  Four values of R$_{\rm min}$ are explored in the 
grid -- 3, 7, 11, and 15$\rstar$.  We explore values of optical depth at 
10.0$\mum$ wavelength, $\tau_{\rm 10}$, between 10$^{-4}$ and 26.  The 
mass loss is assumed to be constant over time, and the dust properties 
assumed are those used for the models of HV 5715 and SSTSAGE052206 
reported by \citet{sarg10}.  This initial grid was computed by running the 
{\bf 2D}ust models on the Royal computer cluster at STScI.  We scale 
our models, as described in \S2.2, to create the final grid of 68600 
models from the initial grid.  For 
each combination of T$_{\rm eff}$, R$_{\rm min}$, and $\tau_{\rm 10}$ 
in the initial grid, the full grid has 56 values of luminosity ranging from 
10$^{3}$--10$^{6} \lsun$ with finer gridding in luminosity for luminosities 
between 10$^{3}$--2$\times$10$^{4} \lsun$.

From this final grid, colors and magnitudes are synthesized from the model 
spectra and compared to the observed colors and magnitudes of O-rich 
AGB, RSG, and extreme AGB candidates in the SAGE sample and of a 
small sample of OH/IR stars from \citet{marsh04}.  The 
coverage of the observed data by the models is generally very good.  
Almost all of the observed data is consistent with the model grid.  The 
very few data points that are not consistent often are likely misidentifications, 
while a few of these may suggest stars with dust shells with parameters 
somewhat outside the range explored in our grid.  The extreme AGB 
candidate points are covered well in the CMDs but poorly in the CCDs, 
suggesting the composition of their dust typically disagrees with that 
assumed for our model grid (i.e., silicates).  The SAGE-Spec data, when 
overlaid on the CMDs and CCDs, lend further support to these ideas -- our 
model grid typically covers the SAGE-Spec O-rich AGBs quite well over 
wide color and magnitude ranges, and most of the extreme AGB candidates 
are more consistent with C-rich AGBs in the SAGE-Spec sample than O-rich 
AGBs in the SAGE-Spec sample.  Further, our optically thicker models are 
largely consistent with a small sample of OH/IR stars from the study by 
\citet{marsh04}.  We determine lower envelopes to the 
K$_{s}$-[8.0] and [3.6]-[24] colors as a function of K$_{s}$-[3.6] color for which 
extreme AGB star colors could be consistent with the colors of the models 
in our oxygen-rich model grid.  We also wish to note that the close general 
matching of our model grid to the observed data in CMDs and CCDs gives 
further support for the dust properties assumed in modeling O-rich evolved 
stars by \citet{sarg10}.

Finally, our model grid has been tested by finding the best fit model from 
the grid to each of the photometric SEDs of stars modeled by \citet{vl99}, 
\citet{groen09}, and \citet{sarg10}.  The quality of fit of model fluxes to observed 
fluxes is quite good.  The matching between luminosity, $\tau_{\rm 10}$, 
and dust mass-loss rate found for HV 5715 and especially SSTSAGE052206 
by our best-fit model here and by the models obtained by \citet{sarg10} is fairly 
close.  Also, our modeling is found to be consistent with that of M stars by 
\citet{vl99} and fairly consistent with that of M stars by \citet{groen09}.  This 
provides verification of the model grid presented in this paper.

\acknowledgements This work is based on observations made with 
the {\it Spitzer Space Telescope}, which is operated by 
the Jet Propulsion Laboratory, California Institute of Technology 
under NASA contract 1407.  We acknowledge funding from the NAG5-12595 
grant, SAGE-LMC {\it Spitzer} grant 1275598, SAGE-SEEDS {\it Spitzer} 
grant 1310534, and Herschel/HERITAGE grant 1381522.  This publication 
makes use of the Jena-St. Petersburg Database of Optical Constants 
\citep{hen99}.  The authors would also like to thank Kevin Volk for helpful 
comments and discussion.  We wish to thank Peter Hauschildt for his 
assistance with the PHOENIX stellar photosphere models.  The authors 
have made use of the SIMBAD astronomical database and would like to 
thank those responsible for its upkeep.  The authors also would like to 
thank Bernie Shiao at STScI for his hard work on the SAGE database and 
his kind assistance.

\clearpage

\begin{figure}[t] 
 \hspace*{-4em}
 \includegraphics[scale=0.7, angle=180]{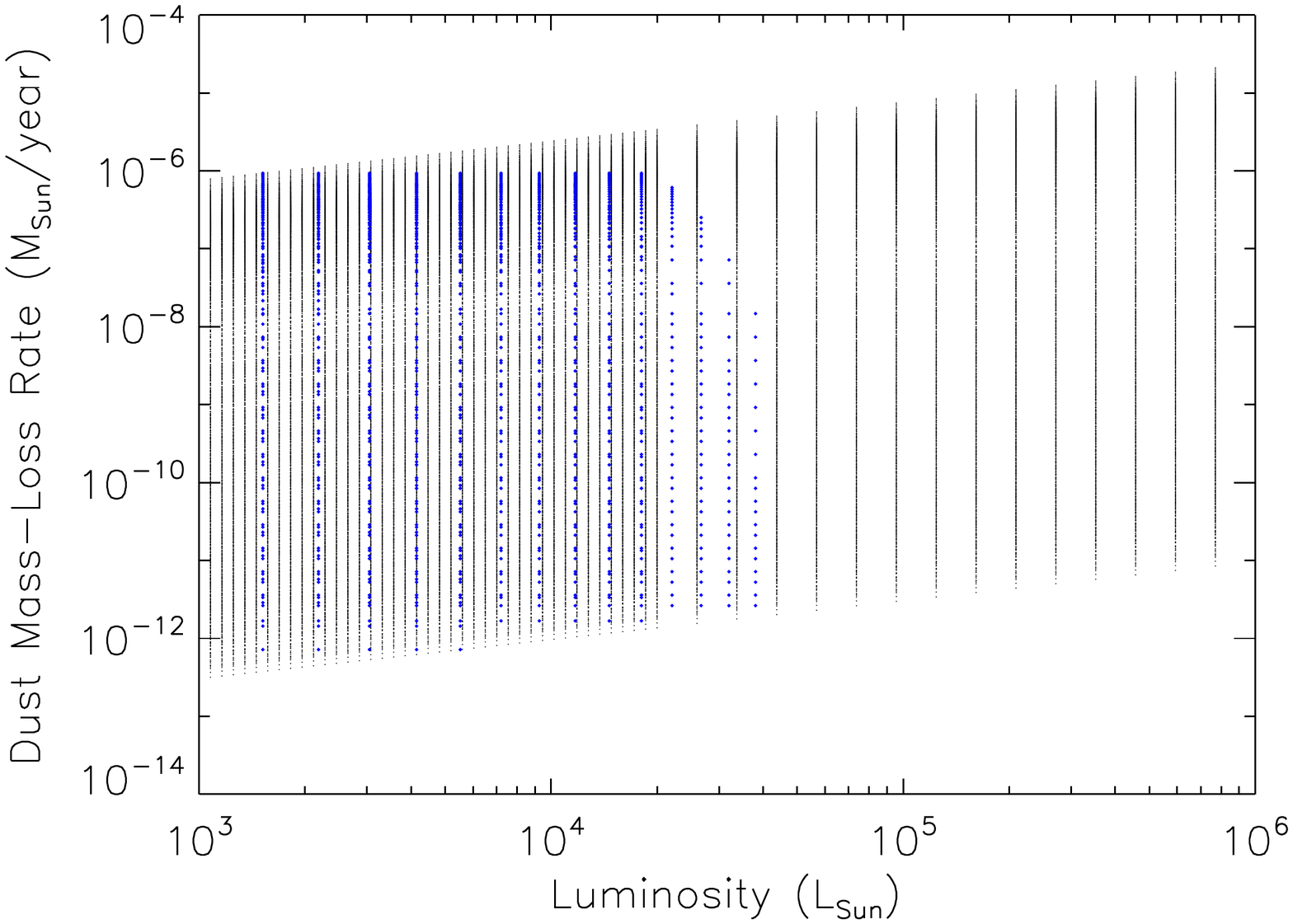}
 \caption[mdotlum]{$\mdot_{dust}$ versus stellar luminosity for 
 the models in our initial and final grids.  The points corresponding 
 to the initial grid are blue, and the points corresponding to the 
 final grid are black.  The $\mdot_{dust}$ is in $\msunyr$, and 
 the luminosities are in $\lsun$.}
\end{figure}

\clearpage

\begin{figure}[t] 
 \hspace*{-8em}
 \includegraphics[scale=0.8, angle=180]{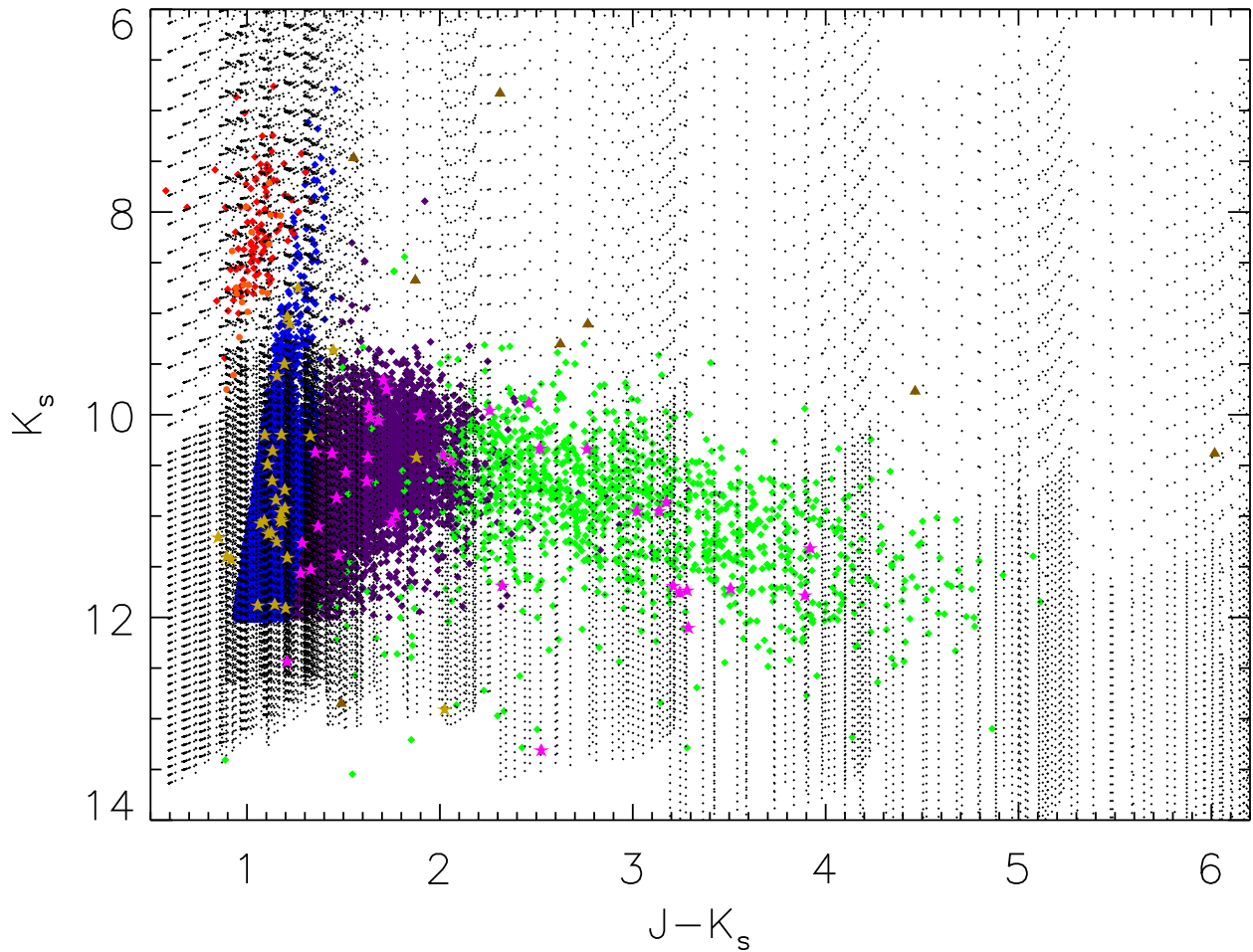}
 \caption[cmd1]{K$_{s}$ versus J-K$_{s}$ color-magnitude diagram.  Of the 
observed sources in the SAGE sample, blue diamonds are 
oxygen-rich AGB candidates, red diamonds are RSG candidates, 
purple diamonds are C-rich AGB candidates, and green diamonds 
are extreme AGB candidates.  The small black points are {\bf 2D}ust 
models from our final model grid (points for the initial grid are not shown 
in this figure or in any of the following figures).  Of the SAGE-Spec sources, 
the O-rich AGB stars are gold star symbols, the RSGs are orange circles, and 
the C-rich AGB stars are magenta star symbols.  OH/IR stars from \citet{marsh04} 
are brown triangles.}
\end{figure}

\clearpage

\begin{figure}[t] 
 \hspace*{-8em}
 \includegraphics[scale=0.8, angle=180]{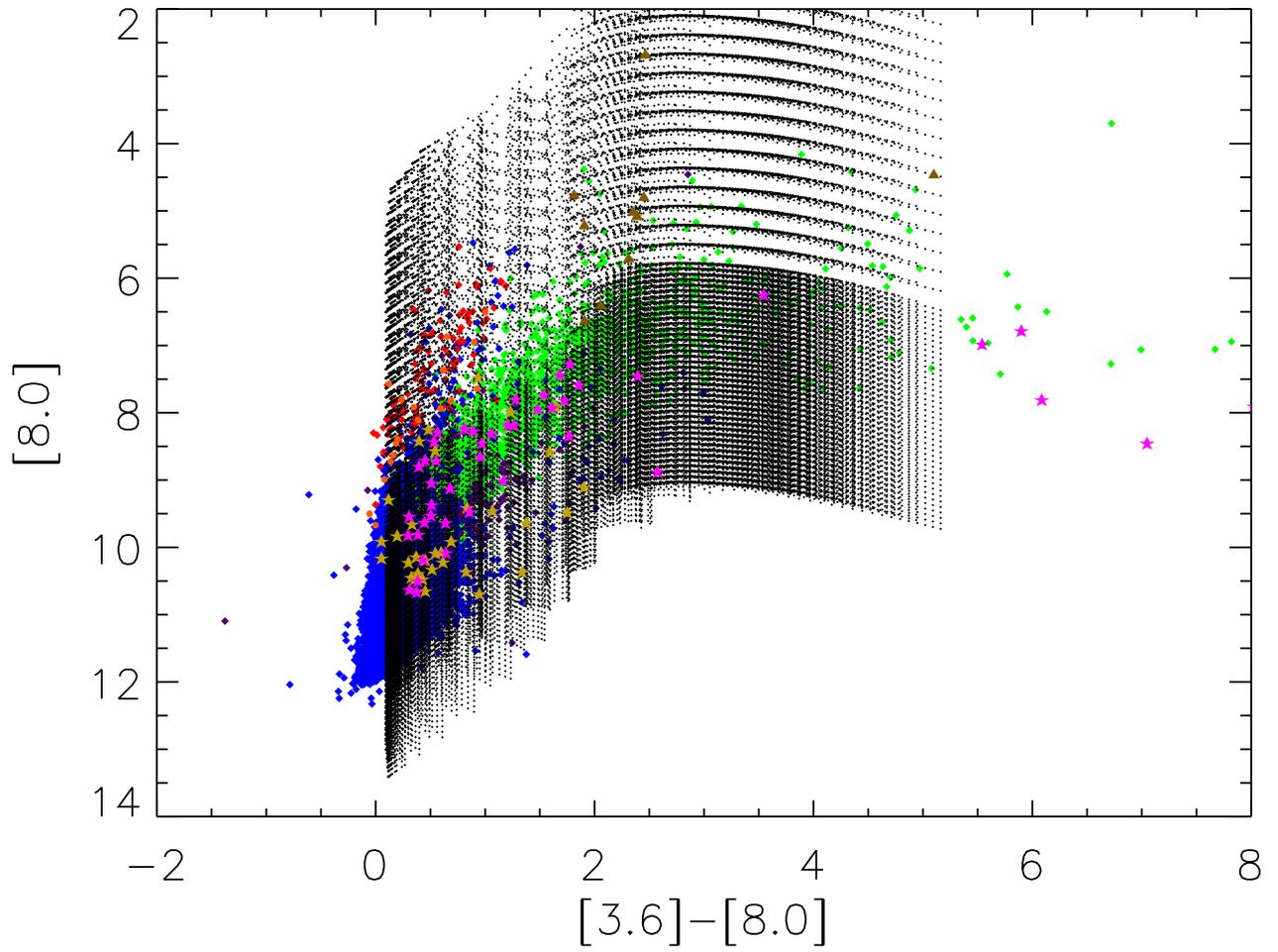}
 \caption[cmd3]{[8.0] versus [3.6]-[8.0] color-magnitude diagram.  
 Same symbol convention as Figure 2.}
\end{figure}

\clearpage

\begin{figure}[t] 
 \hspace*{-8em}
 \includegraphics[scale=0.8, angle=180]{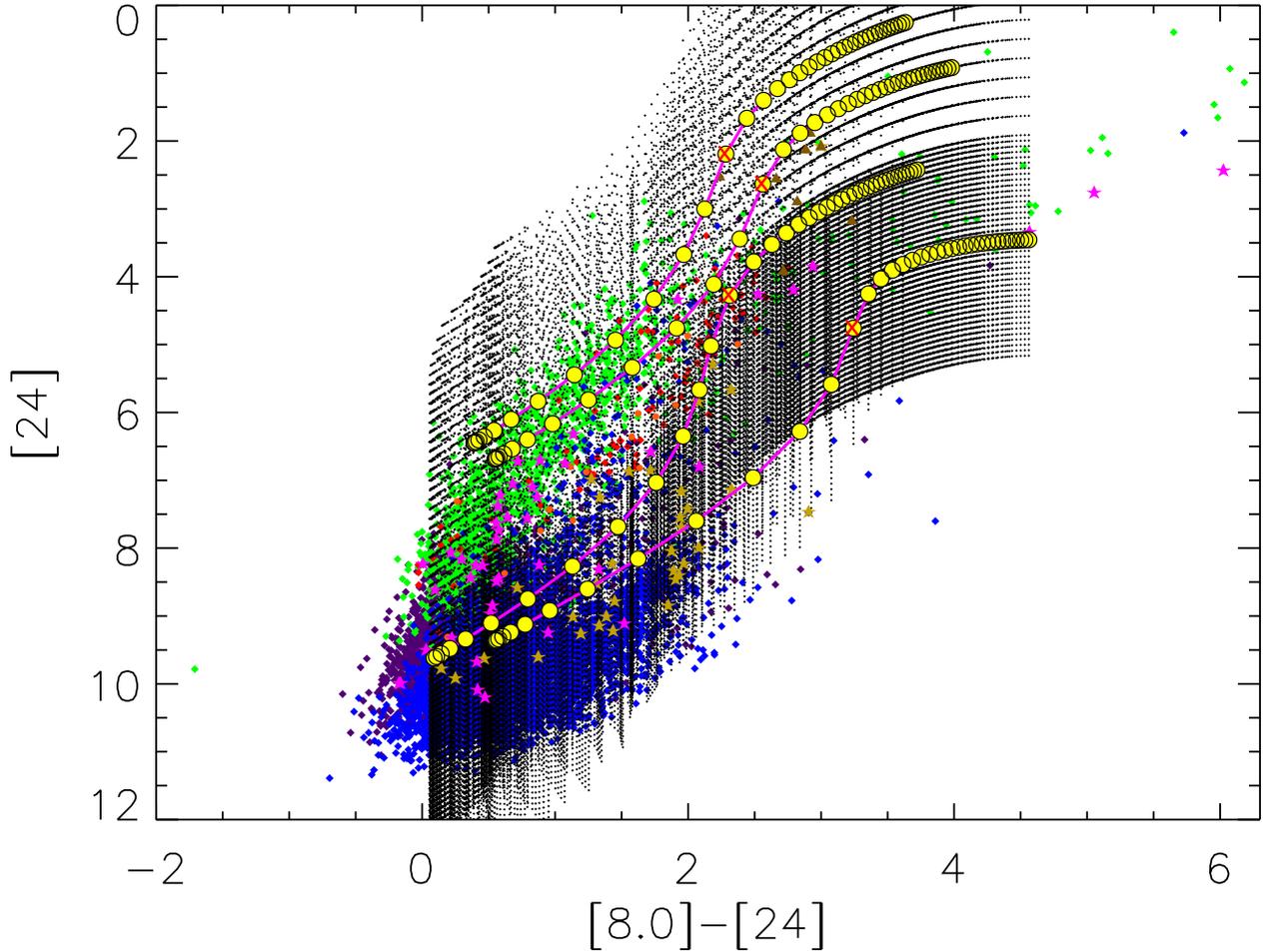}
 \caption[cmd4]{[24] versus [8.0]-[24] color-magnitude diagram.  
 Same symbol convention as Figure 2.  The magenta lines with 
 yellow points overplotted are each a track for a set of models with 
 the same T$_{\rm eff}$, R$_{\rm min}$, and luminosity, with the 
 point corresponding to the model with $\tau_{\rm 10}$  = 1 indicated 
 with a red ``x'' on top of it.  The lowest track (greatest [24] or lowest 
 fluxes) has T$_{\rm eff}$ = 2100\,K, R$_{\rm min}$ = 15$\rstar$, 
 and L = 4800$\lsun$.  The next highest track has T$_{\rm eff}$ = 
 3700\,K, R$_{\rm min}$ = 15$\rstar$,  and L = 16000$\lsun$.  The 
 second highest track has T$_{\rm eff}$ = 2100\,K, R$_{\rm min}$ = 
 7$\rstar$, and L = 57000$\lsun$.  The highest track has T$_{\rm eff}$ 
 = 2700\,K, R$_{\rm min}$ = 7$\rstar$, and L = 120000$\lsun$.}
\end{figure}

\clearpage

\begin{figure}[t] 
 \hspace*{-8em}
 \includegraphics[scale=0.8, angle=180]{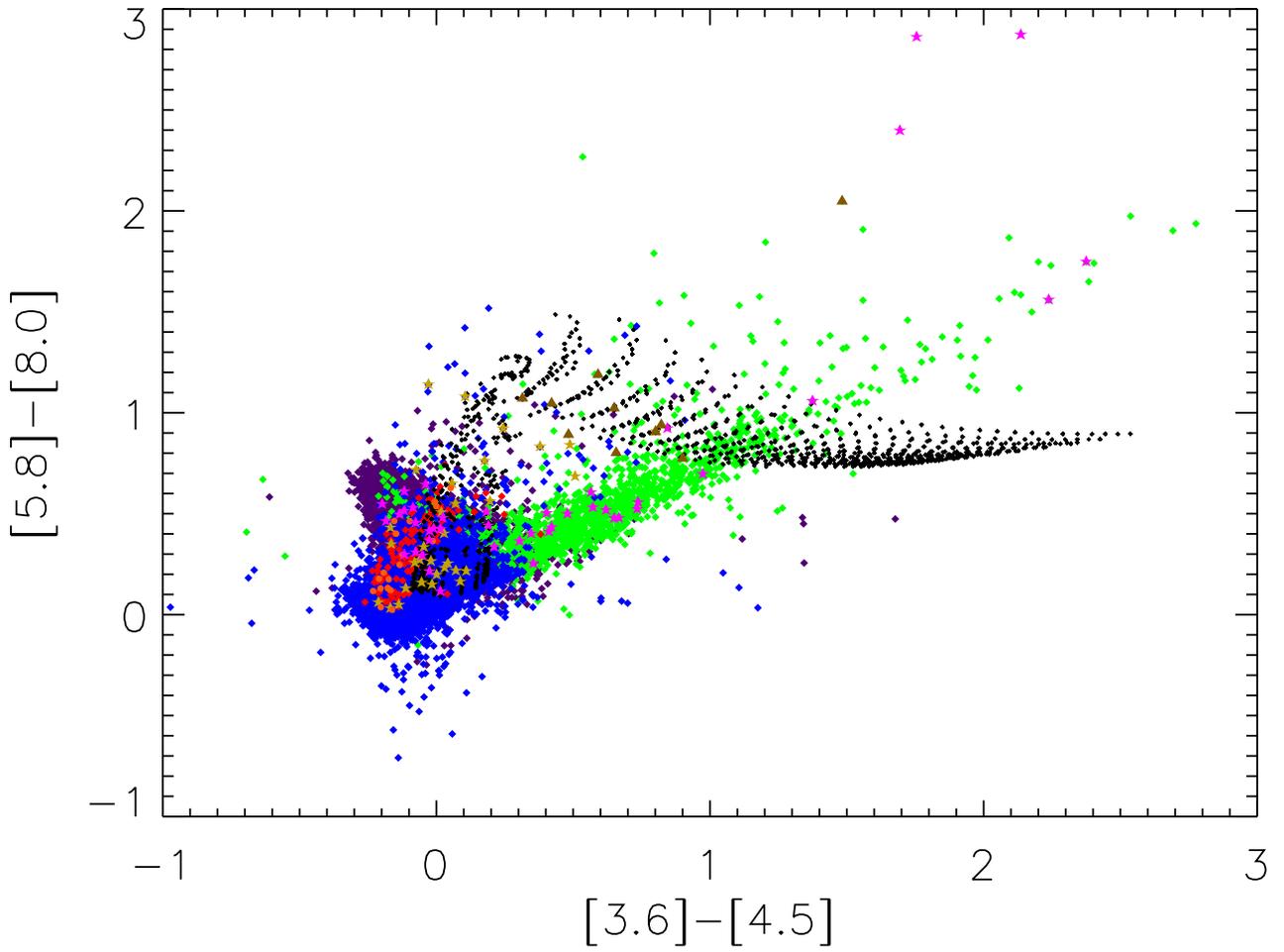}
 \caption[ccd1]{[5.8]-[8.0] versus [3.6]-[4.5] color-color diagram.  
 Same symbol convention as Figure 2.}
\end{figure}

\clearpage

\begin{figure}[t] 
 \hspace*{-8em}
 \includegraphics[scale=0.8, angle=180]{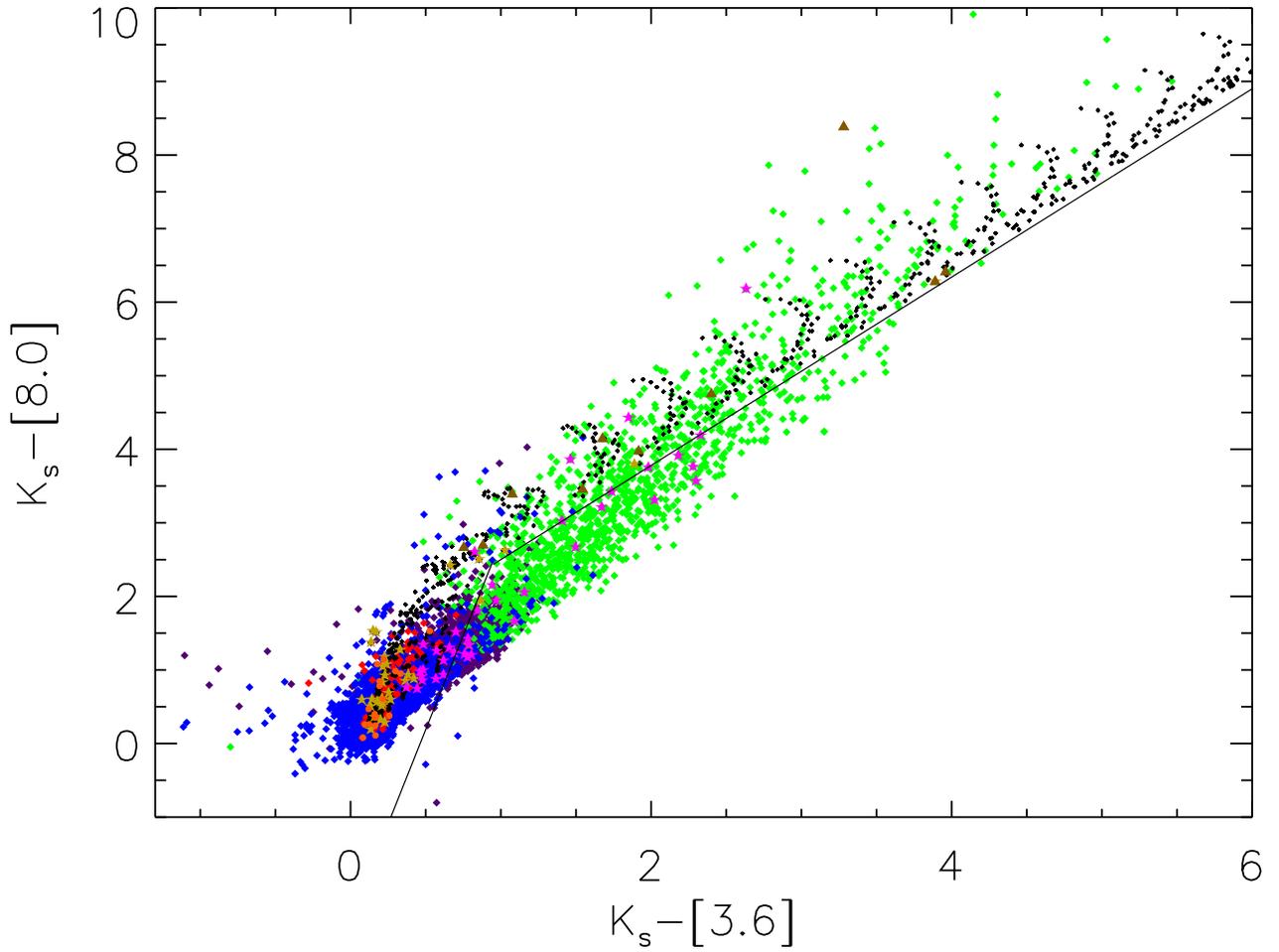}
 \caption[ccd2]{K$_{s}$-[8.0] versus K$_{s}$-[3.6] color-color diagram.  
 Same symbol convention as Figure 2.  The black lines indicate 
 the K$_{s}$-[8.0] colors bluer than which there are no models in our 
 oxygen-rich grid.}
\end{figure}

\clearpage

\begin{figure}[t] 
 \hspace*{-8em}
 \includegraphics[scale=0.8, angle=180]{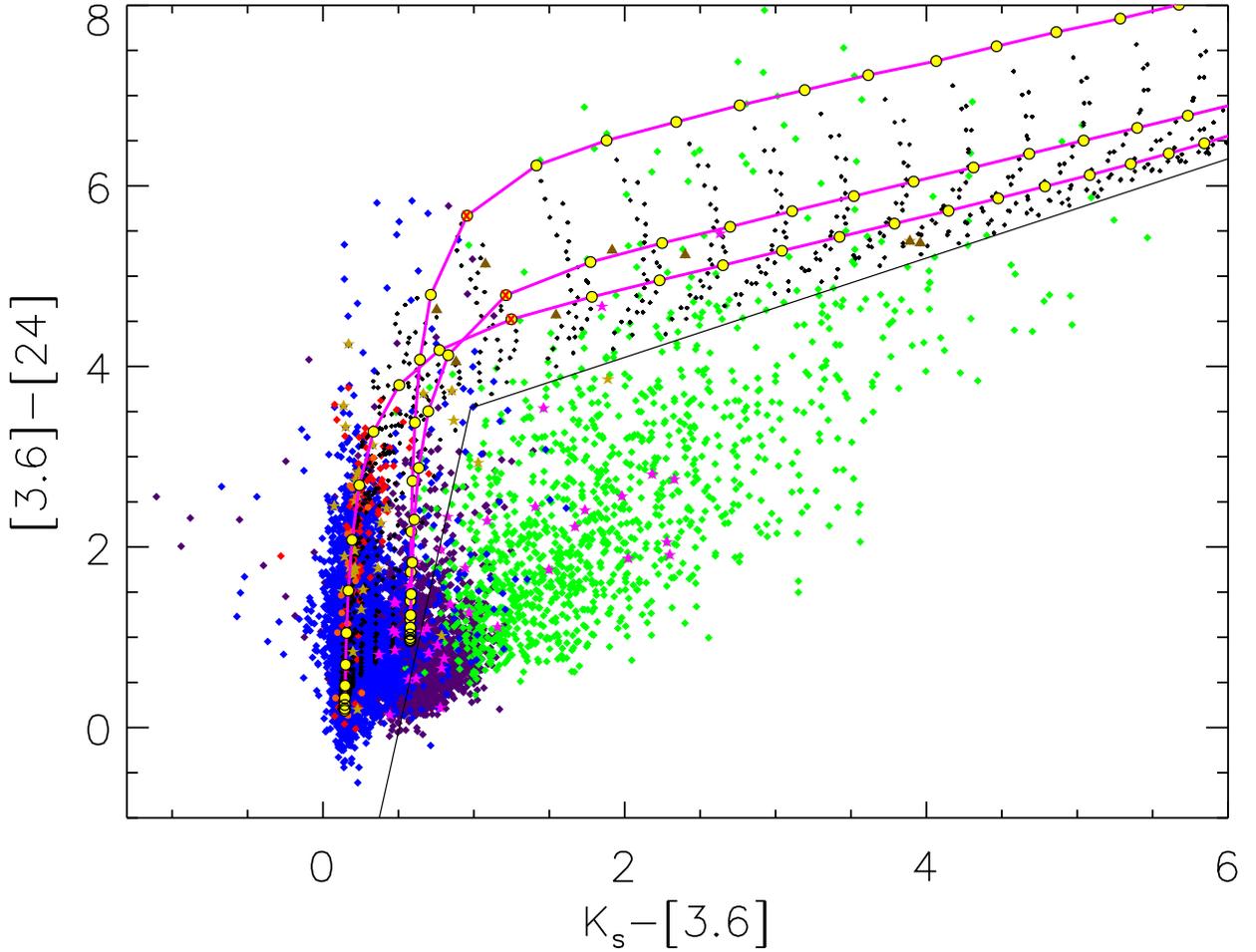}
 \caption[ccd4]{[3.6]-[24] versus K$_{s}$-[3.6] color-color diagram.  
 Same symbol convention as Figure 2.  The black lines indicate 
 the [3.6]-[24] colors bluer than which there are no models in our
 oxygen-rich grid.  The magenta line with yellow points overplotted is 
 a track for a set of models with the same T$_{\rm eff}$, R$_{\rm min}$, 
 and luminosity, with the point corresponding to the model with 
 $\tau_{\rm 10}$  = 1 indicated with a red ``x'' on top of it.  The track 
 beginning at left (lowest K$_{s}$-[3.6] color) has T$_{\rm eff}$ = 3700\,K, 
 R$_{\rm min}$ = 15$\rstar$, and L = 16000$\lsun$.  The track that 
 begins at slightly higher (redder) K$_{s}$-[3.6] color and ends at the 
 greatest [3.6]-[24] colors has T$_{\rm eff}$ = 2100\,K, R$_{\rm min}$ = 
 15$\rstar$,  and L = 4800$\lsun$.  The other track has T$_{\rm eff}$ = 
 2100\,K, R$_{\rm min}$ = 7$\rstar$, and L = 57000$\lsun$.}
\end{figure}

\clearpage

\begin{figure}[t] 
 \includegraphics[scale=0.6]{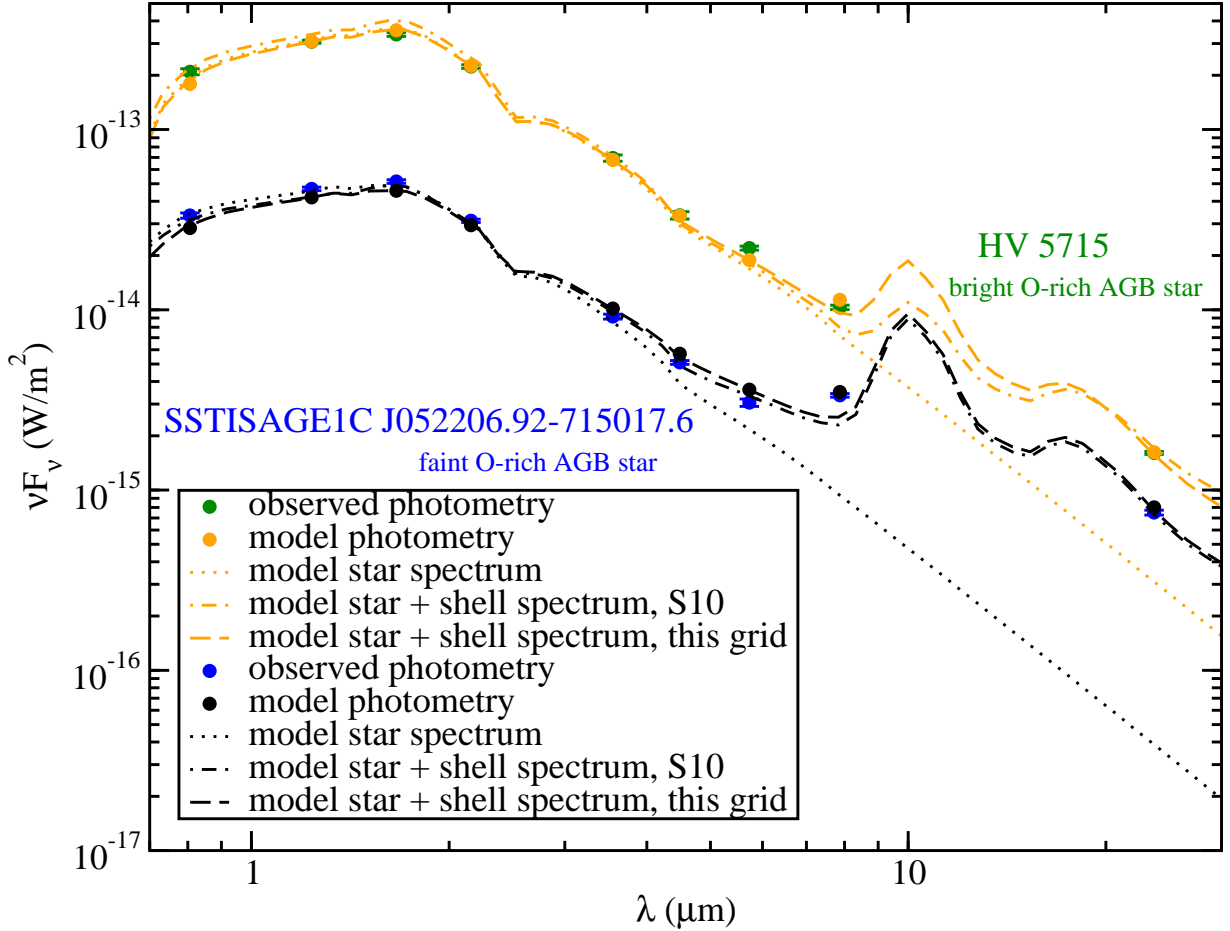}
 \caption[sarg10comp]{Best-fit models from O-rich evolved star grid fit 
 to broadband photometry of each of bright and faint O-rich AGB stars 
 (HV 5715 and SSTISAGE1C J052206.92-715017.6, respectively) fit 
 by \citet{sarg10}.  Green points are observed photometry for bright 
 O-rich AGB star, while orange points and the orange dashed line are 
 the best fit model for this star.  Similarly, blue points are observed 
 photometry for faint O-rich AGB star, and black points and the black 
 dashed line are the best fit model for this star.  For each of the models, 
 large points are the broadband photometry synthesized from the 
 best-fit model, the dashed line is the spectrum of that model, and the 
 dotted line is the spectrum of the stellar photosphere for that model.  
 The two dash-dot lines are the best fits to the bright and faint O-rich 
 AGB stars from \citet{sarg10}, and they are indicated in the legend by 
 ``S10''.}
\end{figure}

\clearpage

\begin{figure}[t] 
 \includegraphics[scale=0.6]{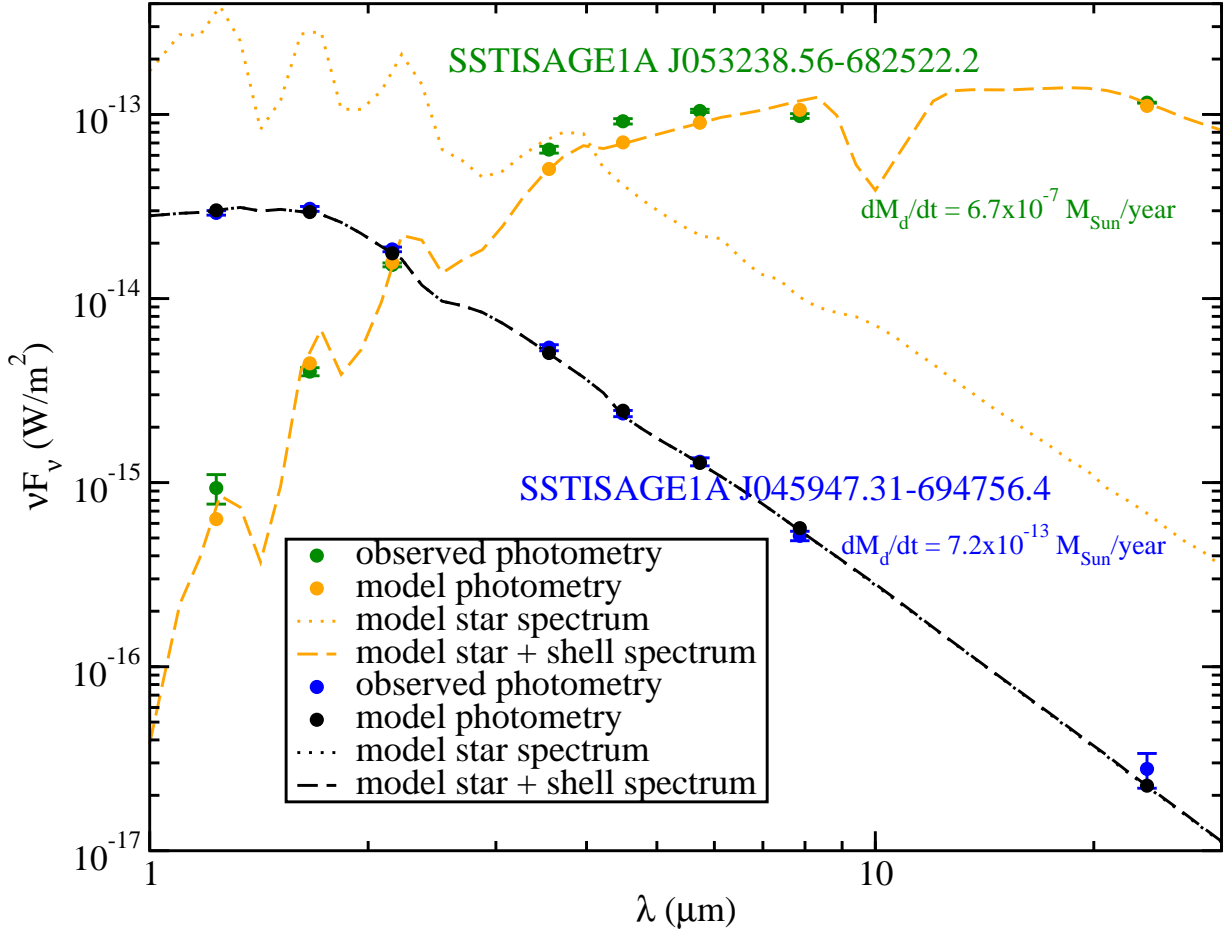}
 \caption[mdotcomp]{Best fit models from O-rich evolved star grid fit 
 to broadband photometry of stars with very high and very low 
 mass-loss rates.  Green points are observed photometry for star with 
 very high mass-loss rate, while orange lines are the best fit model for this star.  
 Similarly, blue points are observed photometry for star with very low 
 mass-loss rate, and black lines are the best fit model for this star.  The 
 same symbol convention as was used for Figure 8 is used here.}
\end{figure}

\clearpage

\begin{figure}[t] 
 \includegraphics[scale=0.6]{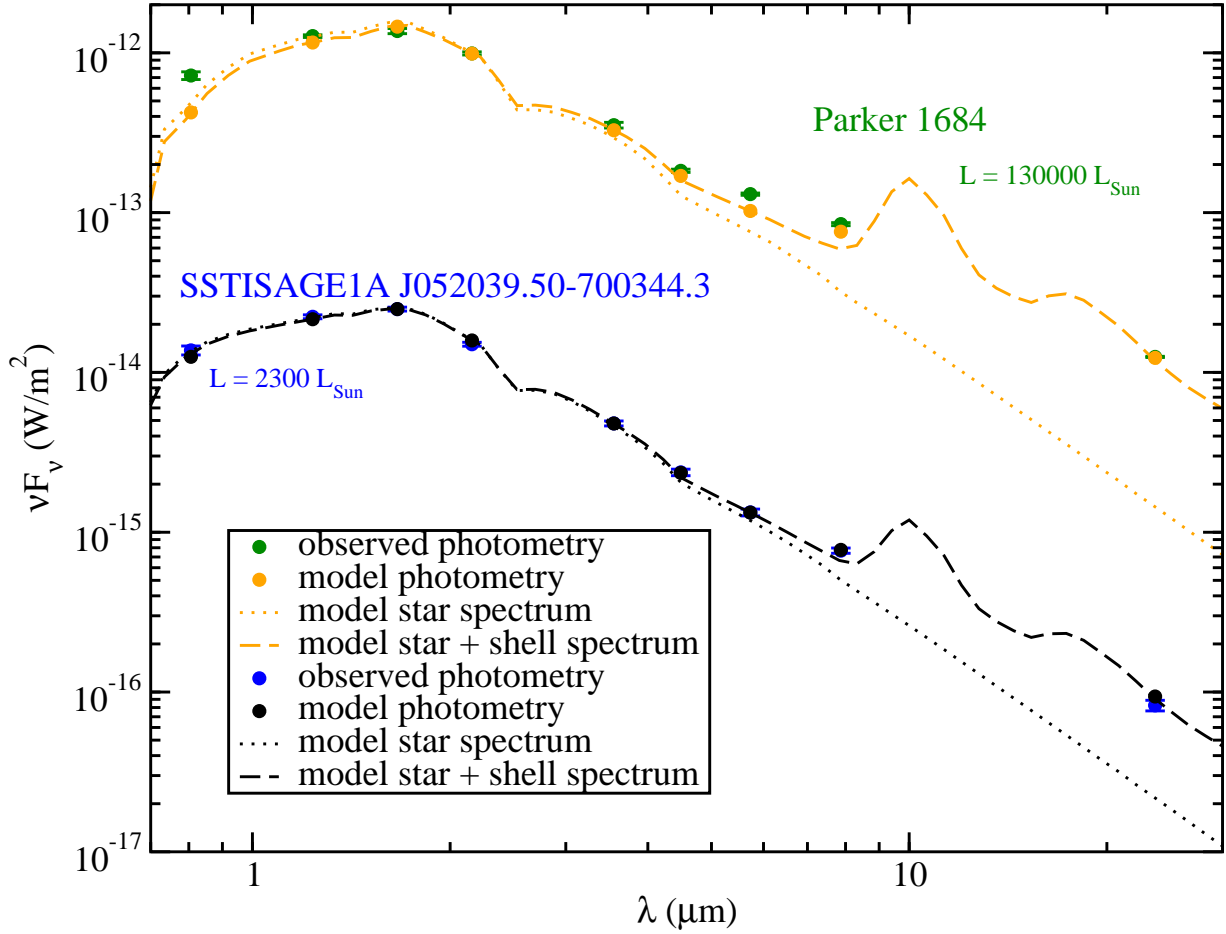}
 \caption[lumcomp]{Best fit models from O-rich evolved star grid fit 
 to broadband photometry of stars with very high and very low 
 luminosities.  Green points are observed photometry for star with 
 very high luminosity, while orange lines are the best fit model for this star.  
 Similarly, blue points are observed photometry for star with very low 
 luminosity, and black lines are the best fit model for this star.  The 
 same symbol convention as was used for Figure 8 is used here.}
\end{figure}

\clearpage

\begin{figure}[t] 
 \includegraphics[scale=0.6]{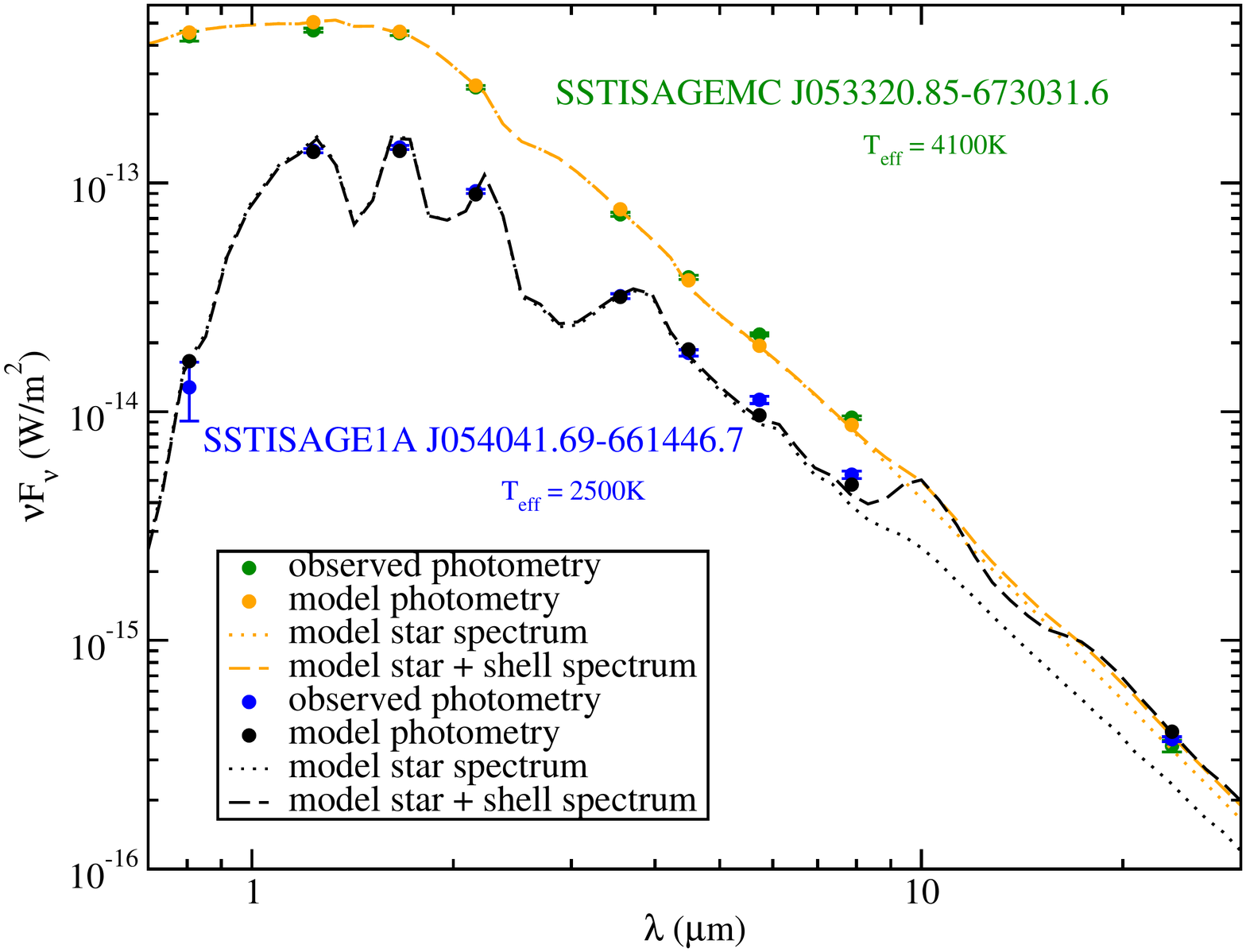}
 \caption[teffcomp]{Best fit models from O-rich evolved star grid fit 
 to broadband photometry of stars with high and low 
 stellar effective temperatures.  Green points are observed photometry for star with 
 high T$_{\rm eff}$, while orange lines are the best fit model for this star.  
 Similarly, blue points are observed photometry for star with low 
 T$_{\rm eff}$, and black lines are the best fit model for this star.  The 
 same symbol convention as was used for Figure 8 is used here.}
\end{figure}

\clearpage

\begin{figure}[t] 
 \includegraphics[scale=0.6]{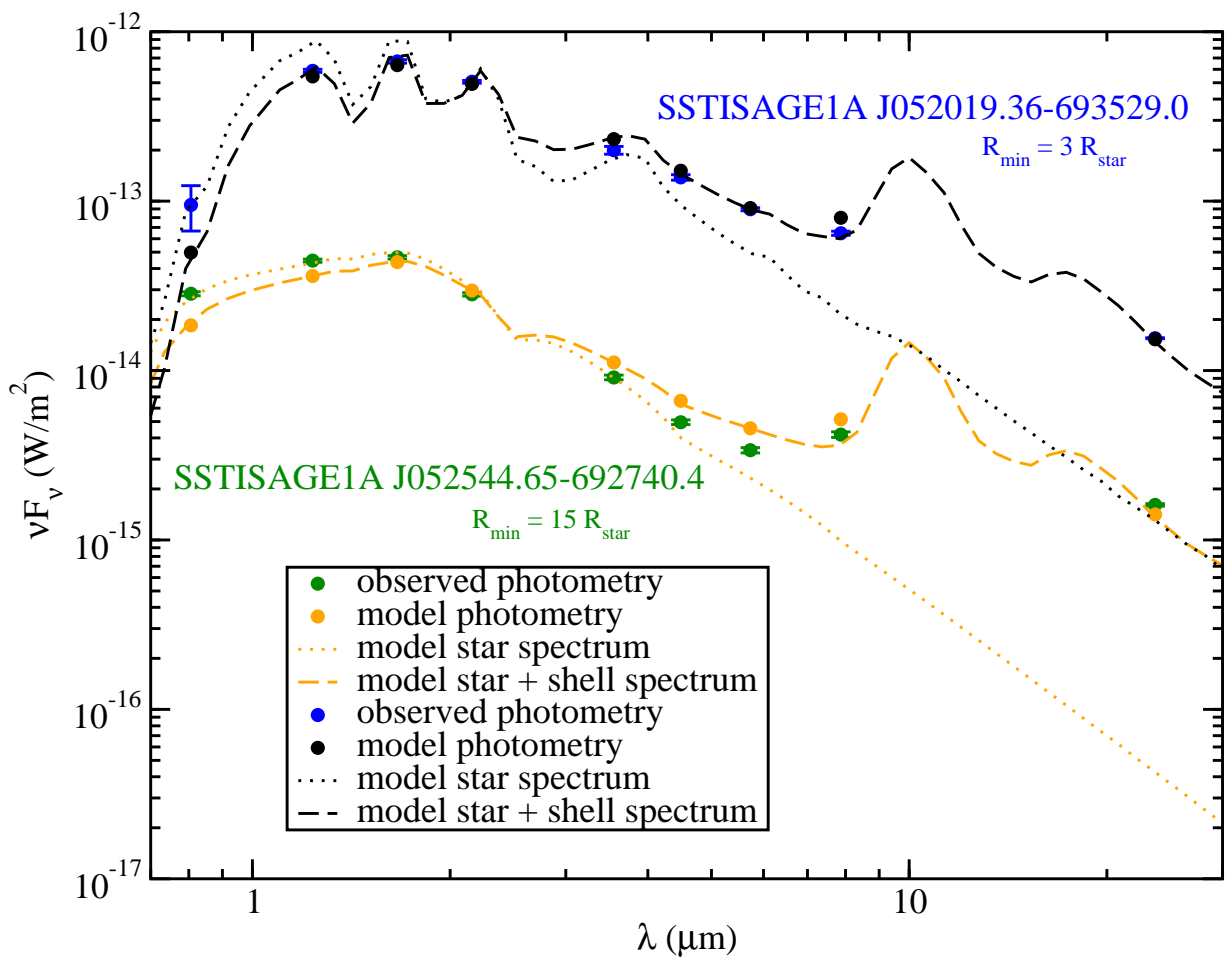}
 \caption[rmincomp]{Best fit models from O-rich evolved star grid fit 
 to broadband photometry of stars with high and low dust shell inner radii.  
 Green points are observed photometry for star with 
 high R$_{\rm min}$, while orange lines are the best fit model for this star.  
 Similarly, blue points are observed photometry for star with low 
 R$_{\rm min}$, and black lines are the best fit model for this star.  The 
 same symbol convention as was used for Figure 8 is used here.}
\end{figure}

\clearpage

\begin{figure}[t] 
 \includegraphics[scale=0.6]{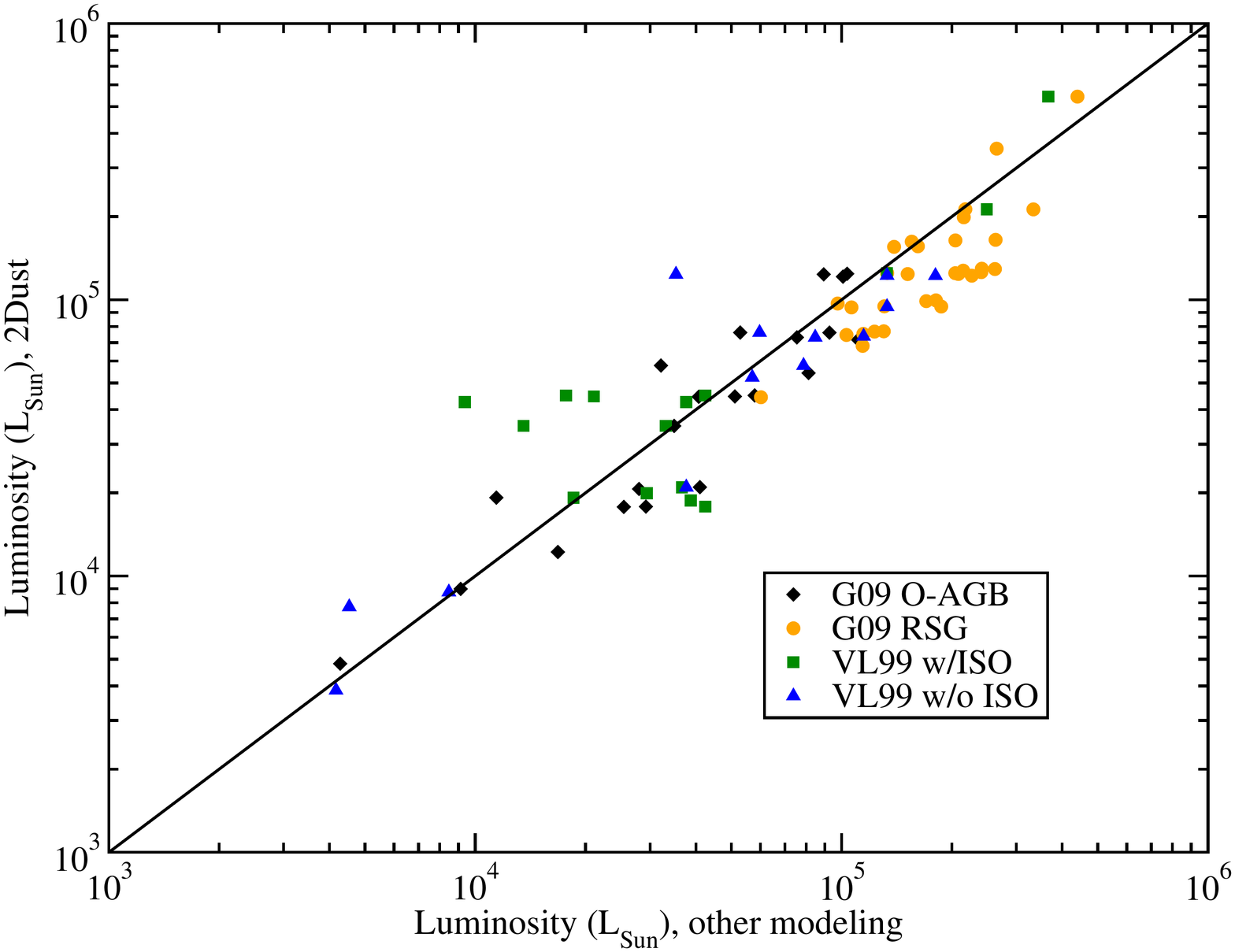}
 \caption[lumcomp]{A comparison of luminosities obtained from fitting {\bf 
 2D}ust models from the GRAMS grid to SAGE data for stars that were 
 also fit by \citet{vl99} and \citet{groen09}.  Black diamonds are for O-rich 
 AGB stars from the \citet{groen09} sample, orange circles are for RSG 
 stars from the \citet{groen09} sample, green squares are for M stars from 
 \citet{vl99} that have ISO spectra, and blue triangles are for M stars from 
 \citet{vl99} that do not have ISO spectra.  The black diagonal line would 
 represent where points would lie if the luminosities obtained by {\bf 
 2D}ust agreed perfectly with those obtained from the other two modeling 
 studies.}
\end{figure}

\clearpage

\begin{figure}[t] 
 \includegraphics[scale=0.6]{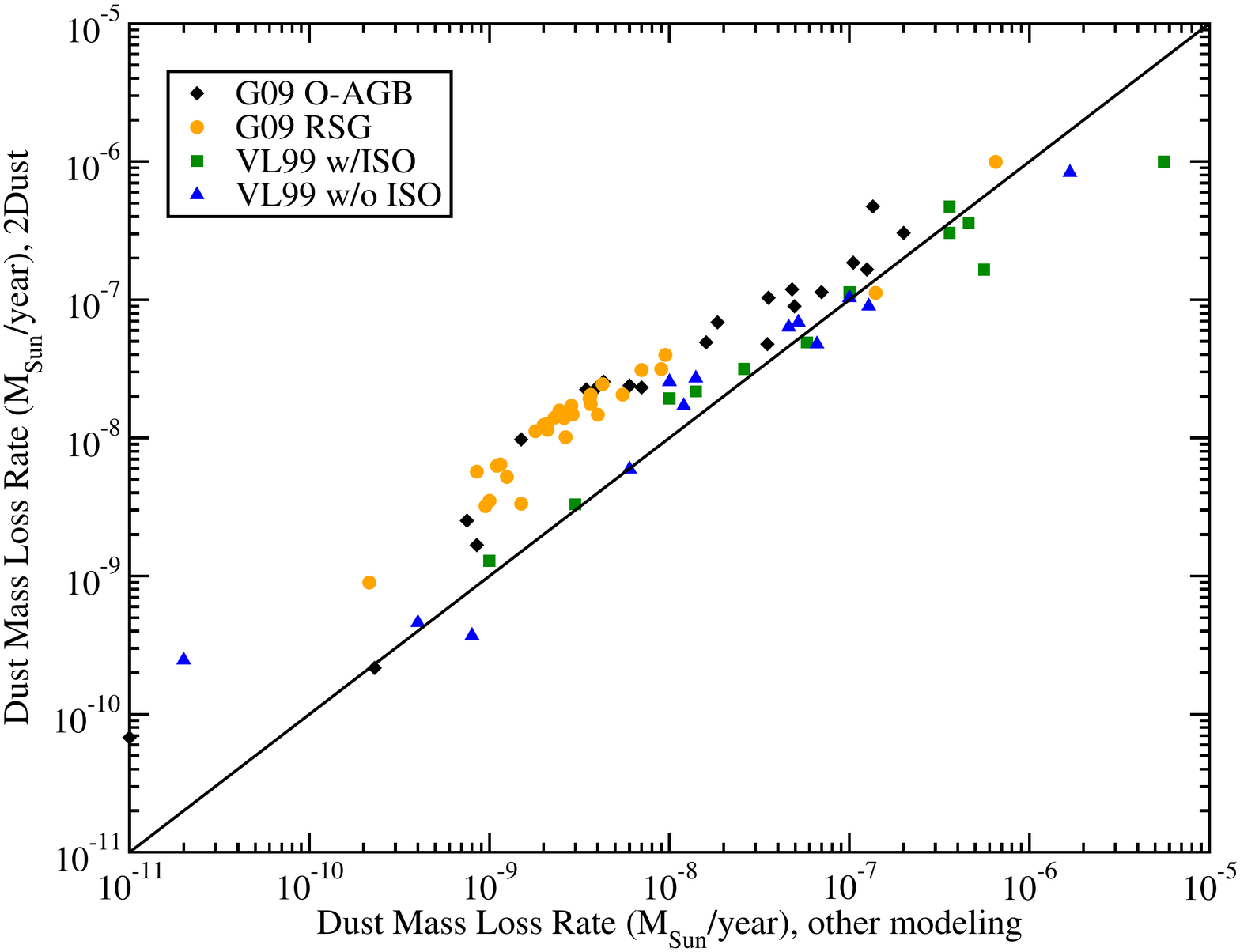}
 \caption[mdotcomp]{A comparison of dust mass-loss rates obtained from 
 fitting {\bf 2D}ust models from the GRAMS grid to SAGE data for stars that 
 were also fit by \citet{vl99} and \citet{groen09}.  Total mass-loss rates 
 from the \citet{vl99} and \citet{groen09} studies were converted to dust 
 mass-loss rates using the gas-to-dust mass ratios that each reference 
 quotes (500 and 200, respectively).  Same symbol convention as for 
 Figure 13, except that the black diagonal line would represent perfect agreement 
 in dust mass-loss rate between {\bf 2D}ust modeling and the modeling 
 reported by the other two studies.}
\end{figure}

\clearpage

\begin{figure}[t] 
 \includegraphics[scale=0.6]{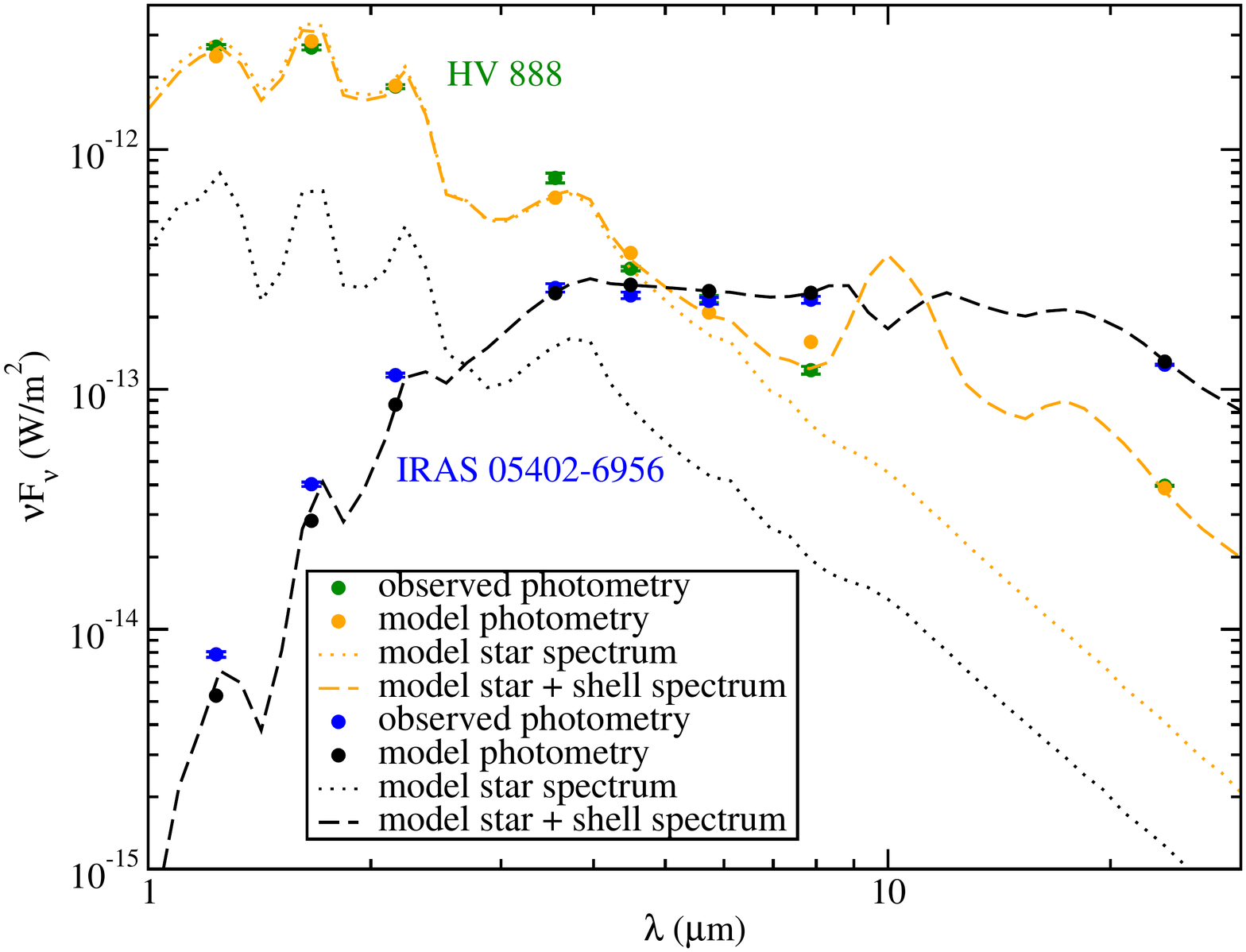}
 \caption[g09vl99comp]{Best fit models from O-rich evolved star grid fit 
 to broadband photometry of stars that were also fit by both \citet{vl99} and 
 \citet{groen09}.  Green points are observed photometry for HV 888, while 
 orange lines are the best fit model for this star.  Similarly, blue points are 
 observed photometry for IRAS 05402-6956, and black lines are the best fit 
 model for this star.  The same symbol convention as was used for Figure 8 
 is used here.}
\end{figure}

\clearpage

\begin{table}[h,t]
{
\caption[parameters]{Parameters for Models in Grid\label{table1}}
\begin{tabular}{lcc}
\hline \hline
	
	& Range of
	& \\
          Parameter
          & Values
          & Increment\\
\hline
{\bf Star} &  & \\
$T_{eff}$ (K) & 2100--4700 & +200\\
Log(g) & -0.5 & \nodata\\
Log($Z/Z_{sun}$)* & -0.5 & \nodata\\
$L_{star} (\lsun)$ &10$^{3}$--10$^{6}$& $\times$1.08, $\times$1.3$^a$\\
{\bf Dust Grains} &  & \\
$\rho_{dust}$ (g/cm$^{3}$)* & 3.3 & \nodata\\
$\gamma$* & -3.5 & \nodata\\
$a_{min} (\mum)$* & 0.01 & \nodata\\
$a_{0} (\mum)$* & 0.1 & \nodata\\
{\bf Assumed Values} &  & \\
$R_{max}$/R$_{min}$* & 10$^{3}$ & \nodata\\
$v_{exp}$ (km/s)* & 10 & \nodata\\
{\bf Dust Shell} &  & \\
$\tau_{10}$ & 10$^{-4}$--26 & $\times$2, +1$^{b}$\\
$R_{min} (\rstar)$ & 3, 7, 11, \& 15 & +4\\
$\mdot_{dust} (\msunyr)$ & 3$\times$10$^{-13}$--3$\times$10$^{-5}$ & \nodata\\
\hline
\end{tabular}
\tablecomments{\footnotesize A KMH grain size 
distribution n(a) $\propto$ a$^{\gamma}$e$^{-a/a_0}$ \citep{kmh94} 
was used for both models.  An asterisk (*) indicates a parameter was 
fixed.}
\tablenotetext{a}{\footnotesize As described in the text, the gridding in luminosity 
is finer, with multiplicative factors of $\simali$1.08, for luminosities less than 
20000$\lsun$, than for luminosities greater than this, which have multiplicative 
factors of $\simali$1.3.}
\tablenotetext{b}{\footnotesize As described in the text, the gridding in 
$\tau_{\rm 10}$ is in multiplicative factors of 2 between $\tau_{\rm 10}$ of 
10$^{-4}$ and 0.41, and the gridding in $\tau_{\rm 10}$ is linear for 
$\tau_{\rm 10}$ between 1 and 26, with increments of 1.}
}
\end{table}

\begin{table}[h,t]
{
\caption[Spectra]{Comparison of Models\label{table2}}
\begin{tabular}{lcccc}
\hline \hline
          
          & SSTSAGE052206
          & SSTSAGE052206
          & HV 5715
          & HV 5715\\
          Parameter
          & all data$^{a}$
          & $\chi^{2}$-min$^{b}$
          & all data$^{a}$
          & $\chi^{2}$-min$^{b}$\\
\hline
L ($\lsun$) & 5100 & 4900 & 36000 & 33000\\
$\tau_{\rm 10}$ & 0.095 & 0.10 & 0.012 & 0.026\\
$\mdotd$ ($\msunyr$) & 2.0$\times$10$^{-9}$ & 2.1$\times$10$^{-9}$ & 2.3$\times$10$^{-9}$ & 1.5$\times$10$^{-9}$\\
\hline
\end{tabular}
\tablenotetext{a}{\footnotesize These columns have 
parameter values that are for the best-fit models from 
\citet{sarg10} for models fit to the {\it Spitzer}-IRS spectra 
and all available photometry.}
\tablenotetext{b}{\footnotesize These columns have 
parameter values that are for the best-fit models found 
by $\chi^{2}$-minimization fitting of photometry over 9 
bands, as described in \S4.2 of this paper.}
}
\end{table}


\begin{thebibliography}{}

\bibitem[Bernard et al.(2008)]{bern08} Bernard, J.-P., et al.\ 
2008, \aj, 136, 919
\bibitem[Blum et al.(2006)]{blum06} Blum, R.~D., et al.\ 2006, 
\aj, 132, 2034
\bibitem[Bonanos et al.(2009)]{bonan09} Bonanos, A.~Z., et al.\ 
2009, \aj, 138, 1003
\bibitem[Bowen(1988)]{bowen88} Bowen, G.~H.\ 1988, \apj, 329, 
299
\bibitem[Boyer et al.(2010)]{boyer10} Boyer, M.~L., et al.\ 2010, 
\aap, 518, L142
\bibitem[Busso et al.(2007)]{busso07} Busso, M., Guandalini, 
R., Persi, P., Corcione, L., \& Ferrari-Toniolo, M.\ 2007, \aj, 133, 2310
\bibitem[Cioni et al.(2000)]{cioni00} Cioni, M.-R.~L., van der 
Marel, R.~P., Loup, C., \& Habing, H.~J.\ 2000, \aap, 359, 601
\bibitem[Cohen et al.(2003)]{coh03} Cohen, M., Wheaton, 
W.~A., \& Megeath, S.~T.\ 2003, \aj, 126, 1090
\bibitem[Dufour et al.(1982)]{duf82} Dufour, R.~J., Shields, 
G.~A., \& Talbot, R.~J., Jr.\ 1982, \apj, 252, 461
\bibitem[Fazio et al.(2004)]{faz04} Fazio, G.~G., et al.\ 
2004, \apjs, 154, 10
\bibitem[Feast(1999)]{fea99} Feast, M.\ 1999, \pasp, 111, 775
\bibitem[Ferrarotti \& Gail(2006)]{ferrga06} Ferrarotti, A.~S., \& Gail, 
H.-P.\ 2006, \aap, 447, 553
\bibitem[Feuchtinger et al.(1993)]{feuch93} Feuchtinger, M.~U., 
Dorfi, E.~A., \& Hofner, S.\ 1993, \aap, 273, 513
\bibitem[Fleischer et al.(1991)]{flei91} Fleischer, A.~J., Gauger, 
A., \& Sedlmayr, E.\ 1991, \aap, 242, L1
\bibitem[Fleischer et al.(1992)]{flei92} Fleischer, A.~J., Gauger, 
A., \& Sedlmayr, E.\ 1992, \aap, 266, 321
\bibitem[Gail et al.(2009)]{gail09} Gail, H.-P., Zhukovska, 
S.~V., Hoppe, P., \& Trieloff, M.\ 2009, \apj, 698, 1136
\bibitem[Gonz{\'a}lez-L{\'o}pezlira et al.(2010)]{gonzlo10} 
Gonz{\'a}lez-L{\'o}pezlira, R.~A., Bruzual-A., G., Charlot, S., 
Ballesteros-Paredes, J., \& Loinard, L.\ 2010, \mnras, 403, 1213
\bibitem[Groenewegen(2006)]{groen06} Groenewegen, 
M.~A.~T.\ 2006, \aap, 448, 181
\bibitem[Groenewegen et al.(2009)]{groen09} Groenewegen, 
M.~A.~T., Sloan, G.~C., Soszy{\'n}ski, I., \& Petersen, E.~A.\ 2009, 
\aap, 506, 1277
\bibitem[Gruendl \& Chu(2009)]{grchu09} Gruendl, R.~A., \& Chu, 
Y.-H.\ 2009, \apjs, 184, 172
\bibitem[Harrington et al.(1988)]{har88} Harrington, J.~P., 
Monk, D.~J., \& Clegg, R.~E.~S.\ 1988, \mnras, 231, 577 
\bibitem[Henning et al.(1999)]{hen99} Henning, T., Il'In, V.~B., 
Krivova, N.~A., Michel, B., \& Voshchinnikov, N.~V.\ 1999, 
\aaps, 136, 405
\bibitem[Hoefner \& Dorfi(1997)]{hoedo97} Hoefner, S., \& 
Dorfi, E.~A.\ 1997, \aap, 319, 648
\bibitem[Hoefner et al.(1998)]{hoef98} Hoefner, S., Jorgensen, 
U.~G., Loidl, R., \& Aringer, B.\ 1998, \aap, 340, 497
\bibitem[H{\"o}fner(2007)]{hoef07} H{\"o}fner, S.\ 2007, Why 
Galaxies Care About AGB Stars: Their Importance as Actors 
and Probes, 378, 145
\bibitem[H{\"o}fner(2009)]{hoef09} H{\"o}fner, S.\ 2009, 
American Institute of Physics Conference Series, 1094, 872
\bibitem[Houck et al.(2004)]{houck04} Houck, J.~R., et al.\ 
2004, \apjs, 154, 18
\bibitem[Ivezic \& Elitzur(1995)]{ivez95} Ivezic, Z., \& Elitzur, 
M.\ 1995, \apj, 445, 415
\bibitem[Jeong et al.(2003)]{jeo03} Jeong, K.~S., Winters, 
J.~M., Le Bertre, T., \& Sedlmayr, E.\ 2003, \aap, 407, 191
\bibitem[Kessler et 
al.(1996)]{kessler96} Kessler, M.~F., et al.\ 1996, \aap, 315, L27
\bibitem[Kim et al.(1994)]{kmh94} Kim, S.-H., Martin, P.~G., 
\& Hendry, P.~D.\ 1994, \apj, 422, 164
\bibitem[Ku{\v c}inskas et al.(2005)]{kucin05} Ku{\v c}inskas, 
A., Hauschildt, P.~H., Ludwig, H.-G., Brott, I., Vansevi{\v c}ius, 
V., Lindegren, L., Tanab{\'e}, T., \& Allard, F.\ 2005, \aap, 442, 
281
\bibitem[Ku{\v c}inskas et al.(2006)]{kucin06} Ku{\v c}inskas, A., 
Hauschildt, P.~H., Brott, I., Vansevi{\v c}ius, V., Lindegren, L., 
Tanab{\'e}, T., \& Allard, F.\ 2006, \aap, 452, 1021
\bibitem[Marigo et al.(2008)]{marig08} Marigo, P., Girardi, L., 
Bressan, A., Groenewegen, M.~A.~T., Silva, L., \& Granato, 
G.~L.\ 2008, \aap, 482, 883
\bibitem[Marshall et al.(2004)]{marsh04} Marshall, J.~R., van 
Loon, J.~T., Matsuura, M., Wood, P.~R., Zijlstra, A.~A., 
\& Whitelock, P.~A.\ 2004, \mnras, 355, 1348
\bibitem[Massey(2003)]{mas03} Massey, P.\ 2003, \araa, 
41, 15
\bibitem[Matsuura et al.(2009)]{matsu09} Matsuura, M., et al.\ 
2009, \mnras, 396, 918 
\bibitem[McDonald et al.(2009)]{mcdon09} McDonald, I., van 
Loon, J.~T., Decin, L., Boyer, M.~L., Dupree, A.~K., Evans, A., Gehrz, 
R.~D., \& Woodward, C.~E.\ 2009, \mnras, 394, 831
\bibitem[Meixner et al.(2006)]{meix06} Meixner, M., et al.\ 
2006, \aj, 132, 2268
\bibitem[Murakami et al.(2007)]{mura07} Murakami, H., et al.\ 
2007, \pasj, 59, 369
\bibitem[Nikolaev \& Weinberg(2000)]{nw00} Nikolaev, S., \& 
Weinberg, M.~D.\ 2000, \apj, 542, 804
\bibitem[Ossenkopf et al.(1992)]{oss92} Ossenkopf, V., 
Henning, T., \& Mathis, J.~S.\ 1992, \aap, 261, 567 
\bibitem[Posch et al.(2007)]{posch07} Posch, T., Mutschke, H., 
Trieloff, M., \& Henning, T.\ 2007, \apj, 656, 615
\bibitem[Rieke et al.(2004)]{riek04} Rieke, G.~H., et al.\ 
2004, \apjs, 154, 25
\bibitem[Sandage \& Walker(1966)]{sandwal66} Sandage, A., 
\& Walker, M.~F.\ 1966, \apj, 143, 313
\bibitem[Sargent et al.(2010)]{sarg10} Sargent, B.~A., et al.\ 
2010, \apj, 716, 878, ``Paper II''
\bibitem[Schlegel et al.(1998)]{schlegel98} Schlegel, D.~J., 
Finkbeiner, D.~P., \& Davis, M.\ 1998, \apj, 500, 525
\bibitem[Skrutskie et al.(2006)]{skrut06} Skrutskie, M.~F., et 
al.\ 2006, \aj, 131, 1163 
\bibitem[Srinivasan et al.(2009)]{srin09} Srinivasan, S., et 
al.\ 2009, \aj, 137, 4810, ``Paper I''
\bibitem[Srinivasan et al.(2010)]{srin10} Srinivasan, S., et 
al.\ 2010, arXiv:1009.2681, ``Paper III''
\bibitem[Ueta \& Meixner(2003)]{um03} Ueta, T., \& Meixner, M.\ 
2003, \apj, 586, 1338
\bibitem[van Loon et al.(1999)]{vl99} van Loon, 
J.~Th., Groenewegen, M.~A.~T., de Koter, A., Trams, N.~R., Waters, 
L.~B.~F.~M., Zijlstra, A.~A., Whitelock, P.~A., \& Loup, C.\ 1999, 
\aap, 351, 559
\bibitem[Verhoelst et al.(2009)]{verhoe09} Verhoelst, T., van 
der Zypen, N., Hony, S., Decin, L., Cami, J., \& Eriksson, 
K.\ 2009, \aap, 498, 127
\bibitem[Volk \& Kwok(1988)]{vokw88} Volk, K., \& Kwok, S.\ 
1988, \apj, 331, 435
\bibitem[Weinberg \& Nikolaev(2001)]{weni01} Weinberg, M.~D., \& 
Nikolaev, S.\ 2001, \apj, 548, 712 
\bibitem[Werner et al.(2004)]{wer04} Werner, M.~W., et al.\ 
2004, \apjs, 154, 1
\bibitem[Whitney et al.(2008)]{whit08} Whitney, B.~A., et al.\ 
2008, \aj, 136, 18
\bibitem[Winters et al.(1997)]{wint97} Winters, J.~M., Fleischer, A.~J., 
Le Bertre, T., \& Sedlmayr, E.\ 1997, \aap, 326, 305
\bibitem[Winters et al.(2000)]{wint00} Winters, J.~M., Le Bertre, T., 
Jeong, K.~S., Helling, C., \& Sedlmayr, E.\ 2000, \aap, 361, 641
\bibitem[Woods et al.(2010)]{woods10} Woods, P.~M., et al.\ 
2010, \mnras, 1801

\end{thebibliography}
\end{document}